\documentclass[final,3p,times]{elsarticle}
\usepackage{multirow}
\usepackage{amsmath}
\usepackage{graphicx}
\usepackage{subcaption}
\usepackage{booktabs}
\usepackage{scalerel}
\usepackage{array}
\usepackage{tikz}
\usepackage{float}

\usepackage{hyperref}

\usepackage{calc}   
\usepackage[dvipsnames]{xcolor}
\usepackage[mathlines]{lineno}
\usepackage{bm}
\usepackage{ragged2e}
\usepackage[most]{tcolorbox}

\newfloat{algorithm}{tbp}{loa}
\floatname{algorithm}{Algorithm}
\journal{Journal of Computational Physics}

\begin{document}

\begin{frontmatter}

\title{LSR-Net: Long-Short-Range Operator Learning for Pattern Dynamics on Manifolds}
\author[inst1]{Qian Serena Hou}
\ead{qhou637@connect.hkust-gz.edu.cn}

\author[inst1,inst2]{Zecheng Gan\corref{cor1}}
\ead{zechenggan@ust.hk}

\cortext[cor1]{Corresponding author}

\affiliation[inst1]{organization={Thrust of Advanced Materials, and Guangzhou Municipal Key Laboratory of Materials Informatics, The Hong Kong University of Science and Technology (Guangzhou)},
                    city={Guangzhou},
                    country={China}}

\affiliation[inst2]{organization={Department of Mathematics, The Hong Kong University of Science and Technology},
                    city={Hong Kong},
                    country={China}}

\begin{abstract}
We propose the \emph{Long-Short-Range Neural Network} (LSR-Net), an extensible operator-learning framework for predicting pattern dynamics on planar domains, spherical surfaces, and general manifolds. 
The method decomposes the forward evolution operator into a long-range component, represented by a compact Fourier multiplier constructed via the \emph{Sum-of-Exponentials} (SOE) approximation, and a short-range component adapted to the underlying geometry and its intrinsic symmetries. 
For general manifolds represented by irregularly sampled point clouds, the long-range component is implemented by Gaussian gridding onto an auxiliary regular grid, where the Fourier multiplier is efficiently applied in $k$-space using FFT and the result is interpolated back to the original sample points.
We evaluate LSR-Net on several benchmark systems, including the Allen--Cahn, Cahn--Hilliard, Schnakenberg, and Turing systems, over planar domains, spherical surfaces, and a blob-shaped manifold. 
Numerical results demonstrate that LSR-Net consistently achieves higher accuracy and improved stability compared with baseline operator-learning models. In particular, for Allen--Cahn dynamics on the sphere, the RMSE is reduced by approximately three orders of magnitude compared with the Spherical Fourier Neural Operator (SFNO). 
Rotation and reflection equivariance tests further confirm that the learned operator is consistent with these geometric transformations. 
These results indicate that LSR-Net provides an effective and robust approach for learning pattern dynamics on complex geometries.

\end{abstract}


\end{frontmatter}

\section{Introduction}
\label{sec:introduction}
Pattern dynamics governed by partial differential equations (PDEs) arise in many physical systems, including phase separation, interfacial motion, and reaction--diffusion processes. 
These models are also central to computational energy science, where phase-field and related PDEs are used to describe microstructure evolution, interfacial transport, and morphology formation in complex materials and multiphase systems~\cite{zhu2018surfactant,cogswell2012coherency}. 
In many practical applications, one seeks fast, accurate, and robust predictions under varying initial conditions, parameters, or geometries, which has motivated growing interest in both AI-assisted surrogate modeling and energy-stable numerical methods~\cite{chan2018uq, shen2018scalar,shen2019new,kou2024energy}. 
Although high-fidelity numerical solvers remain the standard approach, repeated simulation can become prohibitively expensive, especially when the dynamics evolve on curved surfaces or manifolds represented by irregularly sampled point clouds.

These demands have stimulated growing interest in data-driven surrogates for PDE evolution operators. Among these approaches, neural operators aim to learn mappings between function spaces and provide a flexible alternative to solution-by-solution regression~\cite{Lu2021DeepONet, nelsen2021random, li2020neural}. On structured grids, Fourier-based architectures such as the Fourier Neural Operator (FNO)~\cite{Li2021} are particularly effective because they represent nonlocal interactions efficiently in Fourier space. This efficiency, however, relies on regular discretizations and prescribed spectral truncations. As a result, extending Fourier-based operator learning to spherical surfaces and general manifolds represented by point clouds remains challenging, especially in long-horizon prediction settings where both global transport and local effects tied to the underlying geometry must be captured accurately.

Several strategies have been proposed to extend operator learning beyond structured grids, including transformer-based architectures~\cite{Wu2024Transolver,Hao2023GNOT,Zhou2024Unisolver}, graph-based neural operators~\cite{li2023GINO}, convolution-based models on irregular domains~\cite{He2023DeepONet}, and spectral approaches based on learned geometric deformations~\cite{li2023Geo}. 
These developments have substantially broadened the range of admissible geometries. Nevertheless, two challenges remain. First, long-range interactions are still difficult to learn efficiently and accurately on irregular point clouds. Second, local and global components of the evolution operator are often coupled in ways that make it difficult to design architectures that capture both effectively while remaining adaptable across different discretizations and geometries.

In this work, we address these issues through a long-short-range decomposition of the forward evolution operator. We propose the Long-Short-Range Neural Network (LSR-Net), in which long-range interactions are represented by a compact Fourier multiplier constructed from the Sum-of-Exponentials (SOE) approximation~\cite{BeylkinMonzon2005,BeylkinMonzon2007,jiang2008efficient}, while short-range effects are modeled by a local operator adapted to the underlying geometry. 
The SOE representation provides a compact parameterization of nonlocal interactions in Fourier space, and for irregular point clouds on manifolds, it is combined with Gaussian gridding~\cite{dutt1993fast, greengard2004accelerating, barnett2019parallel, peng2023efficient} to transfer the field to an auxiliary Cartesian grid, where the learned Fourier multiplier is applied and the result is interpolated back to the original sample locations.
The short-range component is selected according to the geometry of the data: on planar grids it is implemented by standard convolutions; on the sphere it is paired with DISCO convolutions~\cite{ocampo2022disco}, which are designed to preserve rotational equivariance; and on general manifolds represented by point clouds it is paired with DeltaConv~\cite{wiersma2022deltaconv}, which provides coordinate-independent intrinsic local processing through scalar and tangent-vector features. 
This modular design allows the long-range module to be combined with geometry-specific local operators across different discretizations and geometries.

We evaluate the proposed framework on several benchmark systems for pattern dynamics, including the Allen--Cahn, Cahn--Hilliard, Schnakenberg, and Turing systems posed on planar domains, spherical surfaces, and a blob-shaped manifold. 
The numerical experiments emphasize long-horizon prediction, where error accumulation can significantly degrade pattern fidelity and predictive stability. 
Across all tested scenarios, LSR-Net consistently improves predictive accuracy and stability compared with the baseline operator-learning models considered in this work. 
In particular, for Allen--Cahn dynamics on the sphere, the relative mean-squared error (RMSE) is reduced by approximately three orders of magnitude relative to SFNO~\cite{bonev2023spherical}. 
Rotation and reflection equivariance tests further indicate that the learned operator remains consistent with these geometric transformations. These results demonstrate the effectiveness of LSR-Net for learning pattern dynamics on complex geometries.

The main contributions of this paper are threefold. First, we introduce an SOE-based long-range module that provides a compact Fourier-space representation of nonlocal interactions, can be evaluated efficiently by FFT on regular grids and can be extended to irregular point clouds through Gaussian gridding. Second, we combine this long-range module with geometry-specific short-range operators to obtain a unified operator-learning framework for planar grids, spherical point clouds, and general-manifold point clouds. Third, we demonstrate through numerical experiments and geometric consistency tests that this long-short-range decomposition improves long-horizon predictive accuracy and stability for several representative pattern-forming systems.

The remainder of the paper is organized as follows. Section~2 introduces the SOE-based long-range convolution module and its implementation for irregular point clouds. 
Section~3 presents the LSR-Net framework, its implementations across the three discretization settings, and the training and evaluation procedure. 
Section~4 reports numerical results on planar, spherical, and manifold datasets. 
Section~5 concludes the paper. The appendices contain supplementary experiments, implementation details, summaries of the adopted SR operators, data generation procedures, and model settings.

\section{The SOE-Based Long-Range Convolution for Point Clouds}\label{sec:2}
\begin{figure}[htbp]
	\centering
     \includegraphics[width=0.9\textwidth]{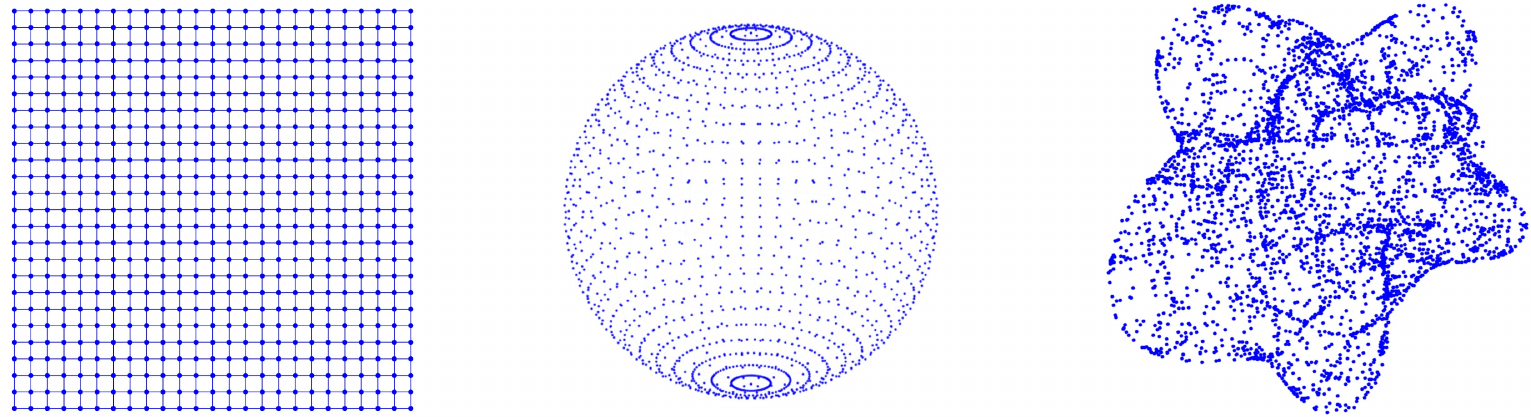}
	\caption{
        \justifying
        {
       Illustrations of the three discretization settings considered in this work: regular grids on a plane, point clouds on the sphere, and point clouds on general manifolds.}
	}
	\label{fig:arch_grid}
\end{figure}

This section presents the long-range (LR) convolution module used in LSR-Net. 
We denote by $\mathcal{K}_{\mathrm{LR}}$ the LR kernel and define the associated LR convolution operator $\mathcal{C}_{\mathrm{LR}}$ by
\begin{equation}
    (\mathcal{C}_{\mathrm{LR}}\phi)(\mathbf{x})
    = \int_{\Omega}
    \mathcal{K}_{\mathrm{LR}}(\mathbf{x}-\mathbf{y})\,
    \phi(\mathbf{y})\, d\mathbf{y}\;,
    \label{eq:phi_continu}
\end{equation}
where $\phi$ denotes the input field. 
In this work, the data are sampled in three discretization settings: regular grids on a plane, point clouds on the sphere, and point clouds on general manifolds, as illustrated in Fig.~\ref{fig:arch_grid}.

The construction of the LR module consists of two steps. 
First, we introduce an SOE-based parameterization of the LR Fourier multiplier. 
Second, to accommodate irregular point clouds, we use Gaussian gridding: the input data are spread onto an auxiliary Cartesian grid, the multiplier is applied there efficiently via FFT, and the result is interpolated back to the sampling locations.
We now describe these two steps in detail.

\subsection{SOE-Based Long-Range Fourier Multiplier}

The Sum-of-Exponentials (SOE) ansatz approximates a broad class of smooth decaying kernels by superpositions of exponential terms. 
Classical approximation results are often stated in terms of real decaying exponentials: after normalizing the distance variable to a bounded interval, for example $r\in[0,1]$, many kernels $\kappa(r)$ can be approximated to a prescribed tolerance $\varepsilon>0$ by a finite sum of decaying exponentials~\cite{devore2009multiscale,BeylkinMonzon2005,BeylkinMonzon2007,jiang2008efficient}:
\begin{equation}
    \bigg| \kappa(r) - \sum_{m=1}^{M} \beta_m e^{-\alpha_m r} \bigg| < \varepsilon,
    \quad \forall r \in [0,1],
\end{equation}
with real coefficients $\beta_m$ and decay rates $\alpha_m>0$. 
Here the interval $[0,1]$ should be understood as a normalized range of pairwise distances. 
This uniform approximation property on a bounded distance range motivates the use of SOE families as compact yet expressive parameterizations of long-range kernels. 
For the learnable LR module used here, however, it is more convenient to work directly with the corresponding complex SOE form, which also captures oscillatory kernel responses~\cite{gao2022kernel,lin2025weighted}. 
Let
\[
    \lambda_i = \alpha_i + \mathrm{i} s_i,
    \qquad \alpha_i > 0,
\]
and pair each $\lambda_i$ with its complex conjugate so that the kernel remains real-valued. 
Accordingly, we parameterize the LR kernel by
\begin{equation}\label{eq:KLR}
    \mathcal{K}_{\mathrm{LR}}(\mathbf{x}-\mathbf{y};\boldsymbol{\theta}_{\mathrm{LR}})
    \approx
    \sum_{i=1}^{P}
    \beta_i \Re\!\left[e^{-\lambda_i |\mathbf{x}-\mathbf{y}|}\right]
    =
    \sum_{i=1}^{P}
    \beta_i \exp\!\left(-\alpha_i\, |\mathbf{x}-\mathbf{y}|\right)
    \cos\!\left(s_i |\mathbf{x}-\mathbf{y}|\right)\;,
\end{equation}
where $\boldsymbol{\theta}_{\mathrm{LR}} = \{\alpha_i, \beta_i, s_i\}_{i=1}^{P}$ are trainable parameters and $P$ is the number of LR convolution channels. 
A key advantage of this form is that each channel admits an analytic 1D Fourier transform:
\[
    \mathcal{F}\!\left[e^{-\alpha |x|}\cos(sx)\right](\xi)
    \propto
    \frac{\alpha}{(\xi+s)^2+\alpha^2}
    +
    \frac{\alpha}{(\xi-s)^2+\alpha^2},
\]
where the proportionality constant can be absorbed into $\beta_i$, with $\xi$ denoting the Fourier frequency variable corresponding to $x$.
This observation leads to an SOE-inspired radial parameterization
of the LR Fourier multiplier in reciprocal space.
Denoting by $\widehat{\phi}(\boldsymbol{\xi})=\mathcal{F}[\phi](\boldsymbol{\xi})$ the Fourier transform of the input field, and letting $\xi = |\boldsymbol{\xi}|$, the convolution theorem yields the Fourier-multiplier representation:
\begin{equation}\label{eq:LR_kernel_FT}
    (\mathcal{C}_{\mathrm{LR}}\phi)(\mathbf{x})
    =
    \mathcal{F}^{-1}\!\left[
        \widehat{\mathcal{K}}_{\mathrm{LR}}(\boldsymbol{\xi})\,
        \widehat{\phi}(\boldsymbol{\xi})
    \right],
    \quad
    \widehat{\mathcal{K}}_{\mathrm{LR}}(\boldsymbol{\xi})
    =
    \sum_{i=1}^{P} \beta_i
    \left[
        \frac{\alpha_i}{(\xi+s_i)^2+\alpha_i^2}
        +
        \frac{\alpha_i}{(\xi-s_i)^2+\alpha_i^2}
    \right].
\end{equation}
This representation is compact, explicit, and can be evaluated directly by FFT on regular grids. 
It is therefore well suited to the repeated evaluations required in neural network training and long-horizon prediction. 
For irregular point clouds, Gaussian gridding is introduced in Sec.~\ref{sec:gaussiangridding} to enable the same LR evaluation.

We finally note that, in the implementation used throughout this work, each LR channel is further modulated by a trainable Tikhonov-type frequency-reweighting factor. 
We write
\begin{equation}
\begin{aligned}
\widehat{\mathcal{K}}_{\mathrm{LR}}^{f}(\boldsymbol{\xi})
&= \sum_{i=1}^P \beta_i 
\left[ 
\frac{\alpha_i}{(\xi + s_i)^2 + \alpha_i^2} 
+ 
\frac{\alpha_i}{(\xi - s_i)^2 + \alpha_i^2} 
\right] w_i(\xi),
\end{aligned}
\label{eq:LR_multiplier}
\end{equation}
where
\begin{equation}
w_i(\xi)=\frac{\xi^{2}}{\xi^{2}+h_i^{2}}\;,
\label{eq:filter_fi}
\end{equation}
and $h_i>0$ is an additional trainable parameter for the $i$-th LR channel. 
The factor $w_i$ suppresses the near-zero modes when $\xi \ll h_i$ and approaches one for $\xi \gg h_i$, thereby reducing low-frequency dominance in the training process without changing the basic SOE structure. 
This simple design is motivated by spectral-bias considerations in neural network training~\cite{rahaman2019spectral,xu2024overview}, and its improvement is examined in~\ref{app:Tik} empirically.
Consequently, each LR channel is compactly described by a complex exponent $\lambda_i=\alpha_i+\mathrm{i}s_i$, a real amplitude $\beta_i$, and a trainable parameter $h_i$, or equivalently by the four real trainable parameters $(\alpha_i,\beta_i,s_i,h_i)$.

\emph{Basic properties.}
The SOE-based LR Fourier multiplier has several useful properties.
Assume that $\alpha_i>0$, $h_i>0$, and $\beta_i\in\mathbb{R}$ for all $i$.
Then $\widehat{\mathcal{K}}_{\mathrm{LR}}^{f}(\boldsymbol{\xi})$ is real-valued and radial, depending only on $\xi=|\boldsymbol{\xi}|$.
Moreover, since $0\leq w_i(\xi)\leq 1$ and
\[
    \frac{\alpha_i}{(\xi\pm s_i)^2+\alpha_i^2}
    \leq
    \frac{1}{\alpha_i},
\]
the multiplier is uniformly bounded by
\begin{equation}
    \left|
    \widehat{\mathcal{K}}_{\mathrm{LR}}^{f}(\boldsymbol{\xi})
    \right|
    \leq
    2\sum_{i=1}^{P}\frac{|\beta_i|}{\alpha_i}.
\end{equation}
Therefore, by Plancherel's theorem, the corresponding LR operator is bounded on $L^2$:
\begin{equation}
    \left\|
    \mathcal{C}_{\mathrm{LR}}\phi
    \right\|_{L^2}
    \leq
    \left(
    2\sum_{i=1}^{P}\frac{|\beta_i|}{\alpha_i}
    \right)
    \|\phi\|_{L^2}.
\end{equation}
Furthermore, in the continuous Euclidean setting, the radial Fourier-multiplier form also implies commutation with translations and rotations. 
Together, these properties support the use of the SOE-based LR module as a stable smooth nonlocal component, with geometry-dependent SR modules to account for local and geometry-specific effects.
Note that for spherical and general manifolds, although this approach does not yield exact discrete equivariance, geometric consistency tests in Sec.~\ref{sec:geom-consistency} show that the resulting equivariance errors remain small in practice.

\subsection{Gaussian Gridding for Irregular Point Clouds}
\label{sec:gaussiangridding}

We now describe how the LR Fourier multiplier is applied when the input is sampled on an irregular point cloud.
Let $\{\mathbf{x}_i\}_{i=1}^{N} \subset [0,L)^d$ be a set of sample locations within a reference box, with associated values $\{\phi_i\}_{i=1}^{N}$.
Here $d$ denotes the dimension of the embedded Euclidean space and of the auxiliary Cartesian grid.
In the settings considered in this work, we have $d=3$ for spherical and general-manifold point clouds embedded in $\mathbb{R}^3$.
The same Gaussian-gridding construction can also be used for planar data with $d=2$.
We restrict our attention to these two low-dimensional cases ($d=2, 3$), where FFT-based Gaussian gridding remains computationally efficient.
Note that Gaussian gridding has been extensively used in non-uniform fast Fourier transform (NUFFT) methods~\cite{dutt1993fast,greengard2004accelerating,barnett2019parallel}; the detailed procedure is presented below.

We begin by representing the discrete data as a weighted sum of Dirac distributions,
\begin{equation}
    \phi(\mathbf{x}) = \sum_{j=1}^{N} \phi_j \, \delta(\mathbf{x} - \mathbf{x}_j),
    \label{eq:phi_dirac}
\end{equation}
which yields the discretized LR response
\begin{equation}
    (\mathcal{C}_{\mathrm{LR}}\phi)(\mathbf{x}_i)
    = \sum_{j=1}^{N}
    \mathcal{K}_{\mathrm{LR}}(\mathbf{x}_i-\mathbf{x}_j)\, \phi_j .
\end{equation}
To regularize the Dirac distributions and make the point-cloud data compatible with FFT, we first convolve the data with the periodized Gaussian kernel
\[
    g_\tau(\mathbf{x}) = \sum_{m \in \mathbb{Z}^d} \exp\!\left(-\frac{\|\mathbf{x}-m L\|^2}{4\tau}\right),
\]
which yields the mollified field
\begin{equation}
    \phi_\tau(\mathbf{x}) = \sum_{j=1}^{N} \phi_j \, g_\tau(\mathbf{x} - \mathbf{x}_j)
\end{equation}
with a width parameter $\tau$. In the computations reported here, we use $\tau = 12\left(\frac{L}{2\pi L_{\mathrm{FFT}}}\right)^2$,
where $L_{\mathrm{FFT}}$ denotes the number of auxiliary regular grid points per spatial direction~\cite{dutt1993fast}. Sampling $\phi_\tau$ on the auxiliary regular grid $\{\mathbf{x}_\ell\}$ gives
\begin{equation}
    \phi_\tau(\mathbf{x}_\ell) = \sum_{j=1}^{N} \phi_j \, g_\tau(\mathbf{x}_\ell - \mathbf{x}_j),
    \qquad \mathbf{x}_\ell \in [0,L)^d.
\end{equation}
Let $N_{\mathrm{FFT}} = L_{\mathrm{FFT}}^d$ denote the total number of auxiliary grid points. Applying the discrete fast Fourier transform $\mathcal{F}_{\mathrm{disc}}$ for $\phi_\tau$ on the regular grid yields
\begin{equation}
    \widehat{\phi}_\tau(\mathbf{k}) = \mathcal{F}_{\mathrm{disc}}\left[ \phi_\tau(\mathbf{x}_\ell)\right] =
    \frac{1}{N_{\mathrm{FFT}}}
    \sum_{\ell \in [0,L_{\mathrm{FFT}}-1]^d}
    \phi_\tau(\mathbf{x}_\ell)\,
    e^{- i \boldsymbol{\xi}_{\mathbf{k}} \cdot \mathbf{x}_\ell},
    \quad \mathbf{k} \in [0,L_{\mathrm{FFT}}-1]^d,
\end{equation}
which can be evaluated with $\mathcal O(N_{\mathrm{FFT}}\log N_{\mathrm{FFT}})$ complexity. 
Here $\mathbf{k}$ denotes the discrete Fourier index, and $ \boldsymbol{\xi}_{\mathbf{k}} = \frac{2\pi}{L}\,\widetilde{\mathbf{k}}$ is the associated wrapped physical wavenumber, where $\widetilde{\mathbf{k}}$ is defined componentwise by
\[
    \widetilde{k}_{\alpha} =
    \begin{cases}
        k_{\alpha}, & 0 \le k_{\alpha} \le L_{\mathrm{FFT}}/2, \\
        k_{\alpha} - L_{\mathrm{FFT}}, & L_{\mathrm{FFT}}/2 < k_{\alpha} \le L_{\mathrm{FFT}}-1,
    \end{cases}
    \qquad \alpha = 1,\dots,d.
\]
Here we assume $L_{\mathrm{FFT}}$ is even, as in all computations reported in this work.
Deconvolving the Gaussian mollifier in Fourier space, we obtain the approximation
\begin{equation}
    \widehat{\phi}(\mathbf{k})
    \approx
    \left(\frac{\pi}{\tau}\right)^{d/2}
    e^{\tau |\boldsymbol{\xi}_{\mathbf{k}}|^2}
    \widehat{\phi}_\tau(\mathbf{k})\;.
\end{equation}
We then apply the SOE-based LR multiplier channel-wise:
\begin{equation}
    \widehat{u}(\mathbf{k})
    =
    \widehat{\mathcal{K}}_{\mathrm{LR}}^{\,f}(\boldsymbol{\xi}_{\mathbf{k}})\,
    \widehat{\phi}(\mathbf{k}),
\end{equation}
and the inverse fast Fourier transform produces the LR response on the auxiliary grid:
\begin{equation}
    u_\tau(\mathbf{x}_\ell)
    =
    \mathcal{F}^{-1}_{\mathrm{disc}}\!\left[\widehat{u}\right](\mathbf{x}_\ell).
\end{equation}
where $\mathcal{F}^{-1}_{\mathrm{disc}}$ denotes the inverse discrete Fourier transform.
Finally, Gaussian interpolation maps this LR response back to the original sample locations, yielding
\begin{equation}
    (\mathcal{C}_{\mathrm{LR}}\phi)(\mathbf{x}_i) \approx u(\mathbf{x}_i)
    =
    \sum_{\ell} u_\tau(\mathbf{x}_\ell) \, g_\tau(\mathbf{x}_i - \mathbf{x}_\ell),
    \qquad i = 1,\dots,N.
\end{equation}

The complete computational procedure for the LR module derived above is summarized in Algorithm~\ref{alg1}.
This computational pipeline provides a unified implementation of the LR module while preserving the compact SOE-based parameterization of the Fourier multiplier. 
In Sec.~\ref{sec:3}, this LR convolution module is paired with geometry-specific SR operators to form the LSR-Net framework. 

\begin{algorithm}[h!]
\caption{SOE-Based Long-Range Convolution Module for Irregular Point Clouds}
\label{alg1}
\centering
\begin{tabular}{p{0.97\linewidth}}
\toprule
\textbf{Input:} Irregular point cloud $\{\mathbf{x}_i\}_{i=1}^N \subset \mathbb{R}^d$, values $\{\phi_i\}_{i=1}^N$, auxiliary-grid parameters $L$, $L_{\mathrm{FFT}}$, and $\tau$, Fourier multiplier $\widehat{\mathcal{K}}_{\mathrm{LR}}^{\,f}$. \\
\textbf{Output:} LR response $\{u(\mathbf{x}_i)\}_{i=1}^N$ \\[4pt]
1. Construct the Dirac representation  
 $\phi(\mathbf{x}) = \sum_{j=1}^N \phi_j \,\delta(\mathbf{x} - \mathbf{x}_j)$. \\[4pt]
2. Spread the point cloud data onto an auxiliary Cartesian grid using the periodized Gaussian kernel $g_\tau$:  
 $\phi_\tau(\mathbf{x}_\ell) = \sum_{j=1}^N \phi_j\, g_\tau(\mathbf{x}_\ell - \mathbf{x}_j)$. \\[4pt]
3. Apply the FFT on the auxiliary grid to obtain $\widehat{\phi}_\tau(\mathbf{k})$ for $\mathbf{k}\in[0,L_{\mathrm{FFT}}-1]^d$. \\[4pt]
4. Construct the wrapped physical wavenumber $\boldsymbol{\xi}_{\mathbf{k}}=(2\pi/L)\widetilde{\mathbf{k}}$ and deconvolve the Gaussian mollifier in Fourier space:  
 $\widehat{\phi}(\mathbf{k}) \approx (\frac{\pi}{\tau})^{d/2} e^{\tau |\boldsymbol{\xi}_{\mathbf{k}}|^2} \widehat{\phi}_\tau(\mathbf{k})$. \\[4pt]
5. Apply the learned long-range Fourier multiplier:  
 $\widehat{u}(\mathbf{k}) = \widehat{\mathcal{K}}_{\mathrm{LR}}^{\,f}(\boldsymbol{\xi}_{\mathbf{k}})\widehat{\phi}(\mathbf{k})$. \\[4pt]
6. Compute the inverse FFT to recover $u_\tau(\mathbf{x}_\ell)$ on the auxiliary grid. \\[4pt]
7. Interpolate the LR response back to the original sample locations:  
 $u(\mathbf{x}_i) = \sum_{\ell} u_\tau(\mathbf{x}_\ell)\, g_\tau(\mathbf{x}_i - \mathbf{x}_\ell)$. \\[4pt]
\bottomrule
\end{tabular}
\end{algorithm}

\section{LSR-Net Framework and Training Procedure}\label{sec:3}
In this section, we combine the long-range (LR) module introduced in Sec.~\ref{sec:2} with geometry-dependent short-range (SR) operators to construct the full LSR-Net framework. This design is motivated by classical kernel-splitting approaches in computational physics, such as Ewald-type decompositions and related particle--mesh methods~\cite{ewald1921ewald,eastwood1980p3m3dp,darden1993particle,hashemi2023computing,gan2025fast, liang2025accelerating,bostrom2026fast,gan2026}, where an interaction kernel is separated into a smooth nonlocal component and a localized correction.
Similar idea has also been extended to AI-surrogate models for first-principle calculations recently~\cite{cheng2025latent,ji2025machine}.
In LSR-Net, the SOE-based LR module provides a compact representation of smooth nonlocal interactions, while the SR module resolves local interfaces and fine-scale structures that depend on the underlying discretization and local geometry.
For regular grids and spherical point clouds, the LR module is evaluated using the Fourier-multiplier/Gaussian-gridding pipeline described in Sec.~\ref{sec:2}; in the regular-grid case, this gridding step is retained for implementation consistency, although a direct FFT realization is also available.
For point clouds on general manifolds, the LR module is applied after first mapping the data onto a latent reference grid through a geometry-informed graph neural operator (GNO)~\cite{li2023GINO}.
We next revisit the operator-learning formulation, then describe the backbone architectures and geometry-specific implementations used in this work.

\subsection{Learning Forward Evolution Operators}

Let $\phi_t=\phi(\bm{x},t)\in\mathcal{X}$ denote the system state at time $t$, where $\mathcal{X}$ is the space of admissible states. 
For a prescribed time increment $T$, the goal is to learn the forward evolution map that advances the state from $\phi_t$ to $\phi_{t+T}$.
Given $N_s$ training pairs $\{(\phi_{t_i}^{(i)},\phi_{t_i+T}^{(i)})\}_{i=1}^{N_s}$ separated by this time increment, we seek a parametric approximation
\[
G_{\bm{\theta}}^{T}:\mathcal{X}\rightarrow\mathcal{X}, \qquad \bm{\theta}\in\Theta,
\]
where $\Theta$ is a finite-dimensional parameter space. 
The learned operator $G_{\bm{\theta}^\star}^{T}$ is expected to approximate the forward map on all admissible states, namely
\[
G_{\bm{\theta}^\star}^{T}[\phi_t]\approx \phi_{t+T}, \qquad \forall\, t\ge 0.
\]
The parameters are obtained by minimizing the mean squared error between the predicted and reference states:
\begin{equation}
\bm{\theta}^{\star}
=
\arg\min_{\bm{\theta}}
\frac{1}{N_s}
\sum_{i=1}^{N_s}
\left\|
G_{\bm{\theta}}^{T}\!\left[\phi_{t_i}^{(i)}\right]
-
\phi_{t_i+T}^{(i)}
\right\|^{2}\;.
\end{equation}
Once this one-step operator has been learned, long-time predictions are generated by autoregressive application of the same map:
\[
\phi_0 \mapsto \phi_T \mapsto \phi_{2T} \mapsto \cdots.
\]

\subsection{Backbone LSR-Net Architectures}

The realization of LSR-Net is organized at two levels. At the operator level, each model combines a long-range operator $\mathcal{C}_{\mathrm{LR}}$ with a geometry-dependent short-range operator $\mathcal{C}_{\mathrm{SR}}$. At the architectural level, this LR--SR decomposition is embedded into a backbone neural network. We first describe the backbone architectures, while the geometry-specific choices of $\mathcal{C}_{\mathrm{SR}}$ and the corresponding data flow are presented in the next subsection.

\begin{figure}[htbp]
    \centering
    \begin{subfigure}[b]{0.48\textwidth}
        \centering
         \includegraphics[width=\textwidth]{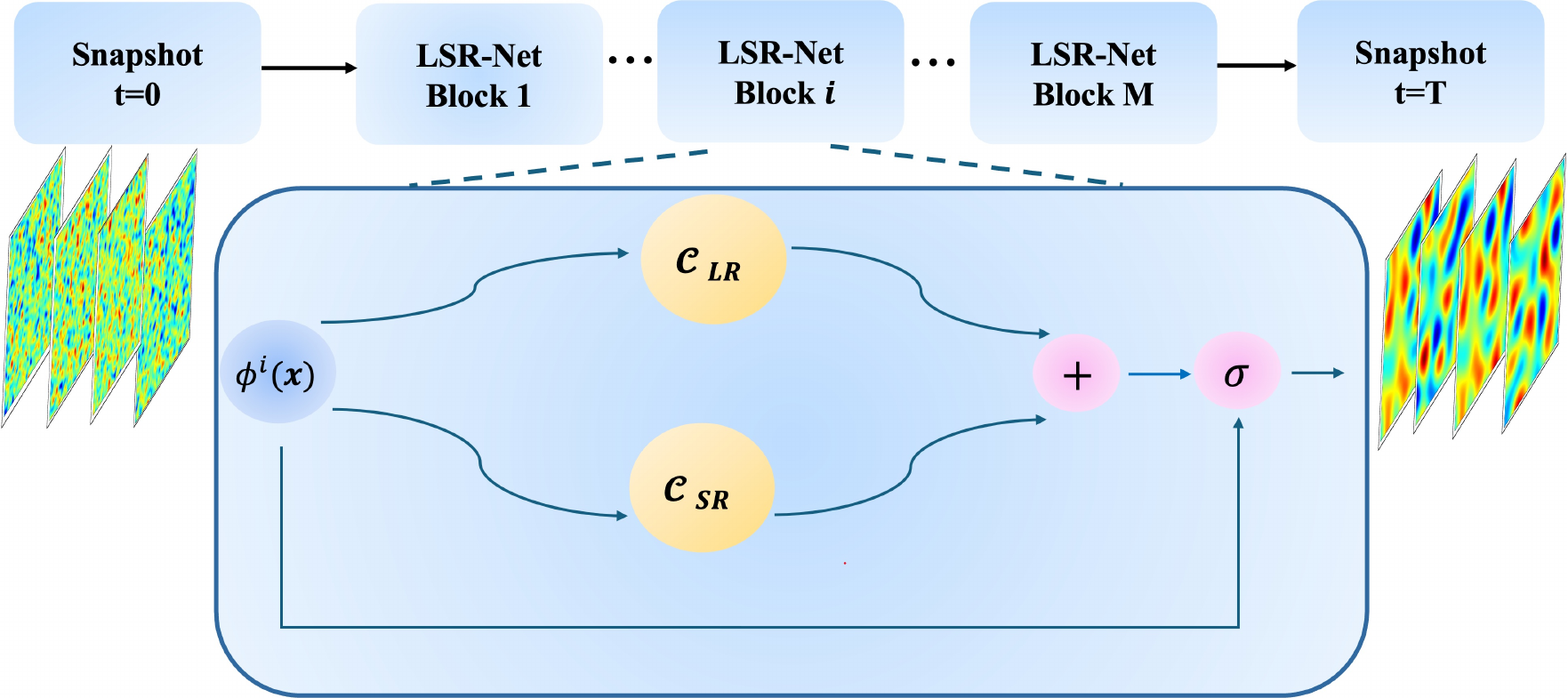}
        \caption{
            \justifying
            The backbone architecture of the base LSR-Net.
            Starting from an input snapshot at $t=0$, the network applies $M$ stacked LSR blocks to predict the snapshot at $t=T$.
            Each block contains a long-range operator $\mathcal{C}_{\mathrm{LR}}$, a short-range operator $\mathcal{C}_{\mathrm{SR}}$, and a nonlinear activation.
            A residual connection is included in each block so that the autoregressive mapping remains close to the identity.
        }
        \label{fig:arch_NN}
    \end{subfigure}
    \hfill
    \begin{subfigure}[b]{0.48\textwidth}
        \centering
        \includegraphics[width=\textwidth]{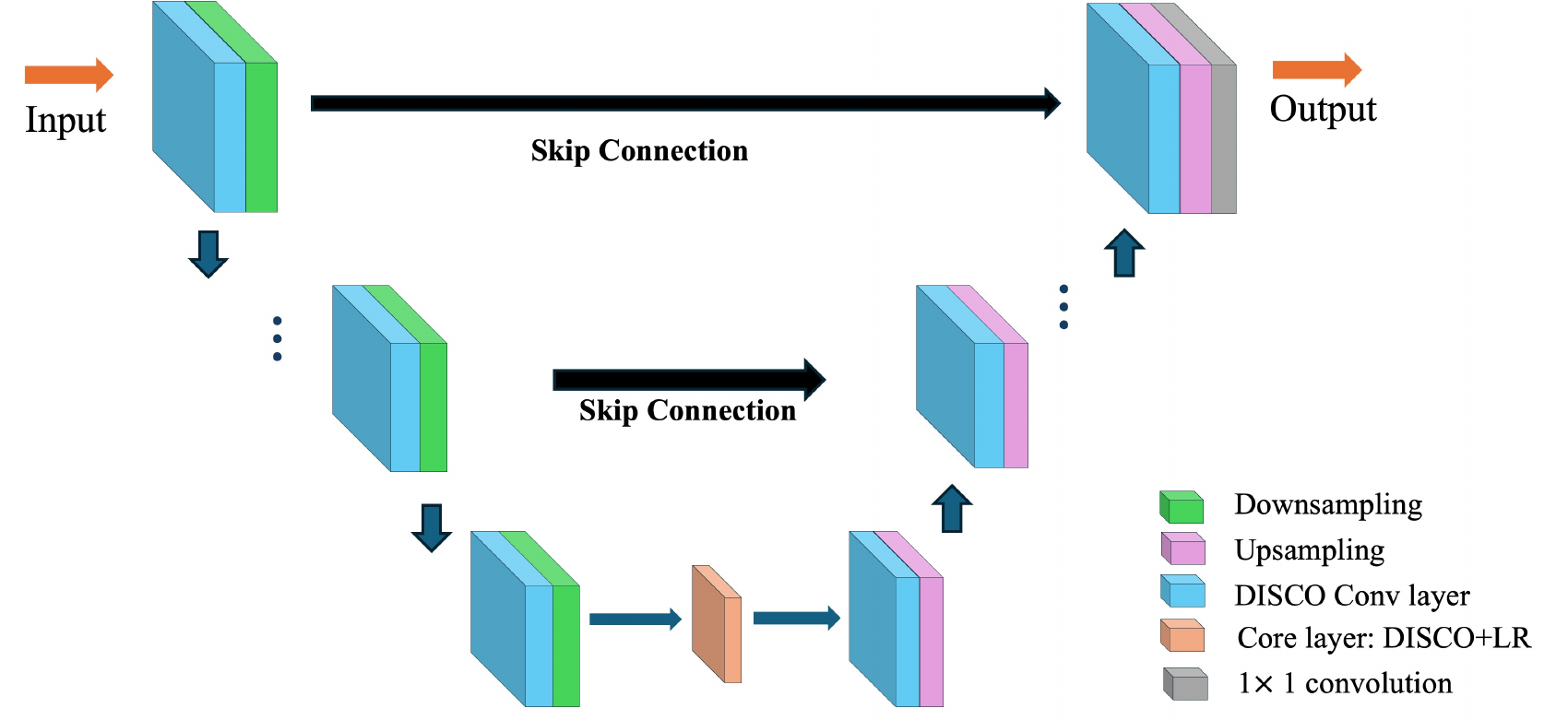}
        \caption{
            \justifying
            The backbone architecture of LSR-U-Net.
            Each LSR-U-Net block consists of an \emph{encoder} with DISCO-based short-range convolution layers~\cite{ocampo2022disco} and downsampling, a \emph{core layer} that combines the long-range and DISCO-based short-range modules, and a \emph{decoder} that upsamples the learned features while concatenating skip connections from the encoder.
        }
        \label{fig:arch_NN_unet}
    \end{subfigure}
    \caption{The two backbone architectures used in this work. Left: the base LSR-Net. Right: the LSR-U-Net.}
\end{figure}
In this work, we consider two backbone architectures. For regular grids on a 2D plane and irregular point clouds on general manifolds, we use the \emph{base LSR-Net} architecture shown in Fig.~\ref{fig:arch_NN}, which consists of $M$ stacked LSR-Net blocks. Each block contains the long-range operator $\mathcal{C}_{\mathrm{LR}}$, the short-range operator $\mathcal{C}_{\mathrm{SR}}$, a pointwise nonlinearity $\sigma$, and a residual connection. This design follows the residual block structure commonly used in neural operator architectures.
For the base LSR-Net, the learned one-step evolution operator can be written as:
\begin{equation}
    G^T_{\bm{\theta}}[\phi]
    =
    \left( \sigma \circ g_{\bm{\theta}_M} \right)
    \circ
    \left( \sigma \circ g_{\bm{\theta}_{M-1}} \right)
    \circ \cdots \circ
    \left( \sigma \circ g_{\bm{\theta}_1} \right)[\phi],
\end{equation}
where each block operator \(g_{\bm{\theta}_i}\) is defined by
\begin{equation}
    g_{\bm{\theta}_i}[\phi^i]
    =
    \phi^i
    +
    \mathcal{C}_{\mathrm{LR}}^{(i)}[\phi^i]
    +
    \mathcal{C}_{\mathrm{SR}}^{(i)}[\phi^i]\;.
    \label{kernel}
\end{equation}
Here, \(\phi^i\) denotes the intermediate feature field in the $i$th block, \(\mathcal{C}_{\mathrm{LR}}^{(i)}\) is the SOE-based long-range operator, \(\mathcal{C}_{\mathrm{SR}}^{(i)}\) is the geometry-dependent short-range operator, and \(\bm{\theta}=\{\bm{\theta}_i\}_{i=1}^M\) collects all trainable parameters. In all numerical experiments reported in this work, the nonlinear activation is fixed to be \texttt{Tanh}.
For spherical data, we embed the same LR--SR block into a U-Net encoder--decoder architecture~\cite{ronneberger2015unet}, yielding the \emph{LSR-U-Net} architecture shown in Fig.~\ref{fig:arch_NN_unet}. 
Each LSR-U-Net block consists of an \emph{encoder} with DISCO-based short-range convolution layers~\cite{ocampo2022disco} and downsampling, a \emph{core layer} that combines the long-range and DISCO-based short-range modules, and a \emph{decoder} that upsamples the learned features while concatenating skip connections from the encoder.
A more detailed description of the LSR-U-Net architecture is provided in~\ref{app:LSR-U-Net}. 
The same LR--SR block can in principle be incorporated into other backbone architectures.

\subsection{Geometry-Specific Implementations of the LR-SR Module}

\begin{figure}[htbp]
	\centering
     \includegraphics[width=1\textwidth]{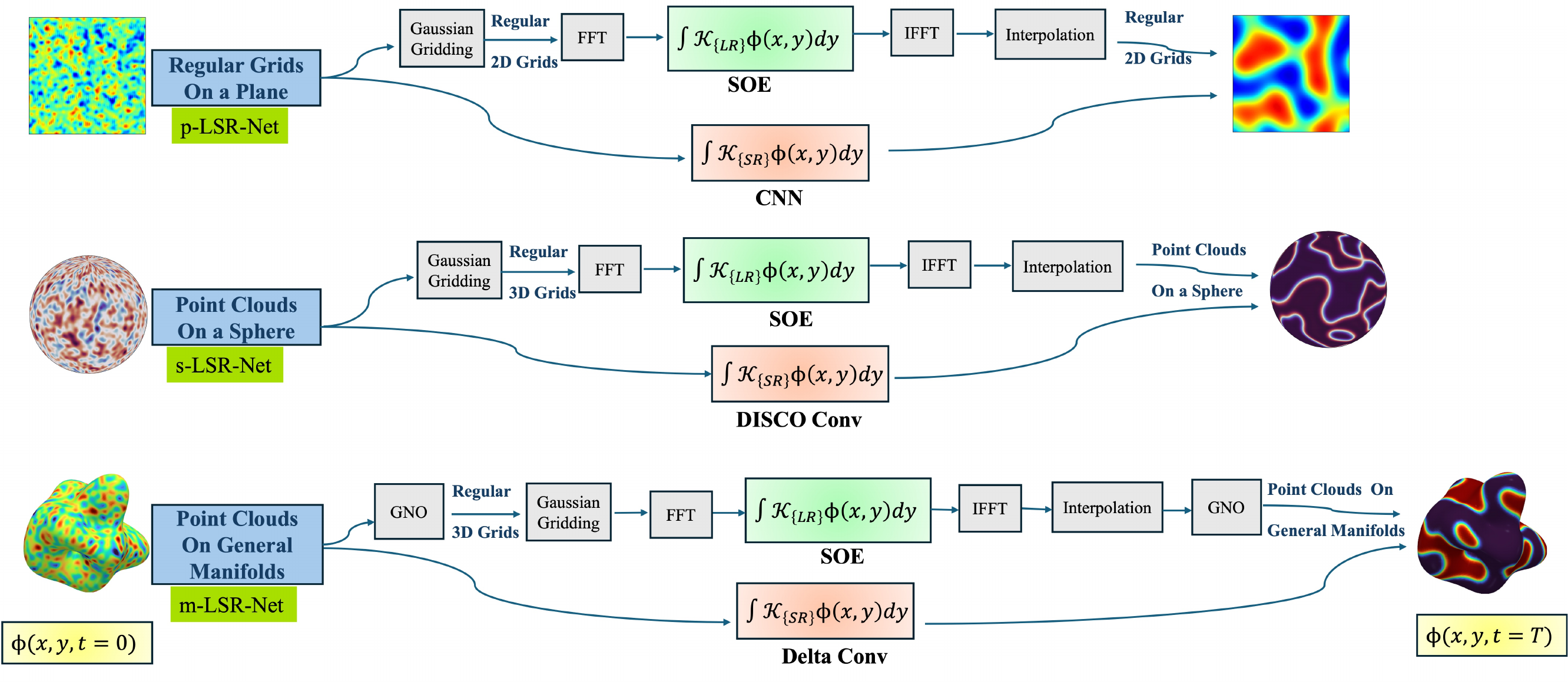}
	\caption{
        \justifying
        Geometry-specific implementations of the LR--SR module.
        (a) For regular grids on a plane, \emph{p-LSR-Net} pairs the SOE-based LR module with standard local convolutions.
        (b) For spherical point clouds, \emph{s-LSR-Net} embeds the LR module in an LSR-U-Net backbone and uses DISCO convolution for the SR component.
        (c) For point clouds on general manifolds, \emph{m-LSR-Net} uses a geometry-informed GNO operator~\cite{li2023GINO} to map features to a latent reference grid in the LR branch, while DeltaConv provides the SR correction directly on the manifold point cloud.
    }
\label{fig:arch_LSR_All}
\end{figure}

We now describe how the LR--SR block is realized for the three data representations considered in this work, as summarized in Fig.~\ref{fig:arch_LSR_All}.
Across these realizations, the LR component follows the SOE-based Fourier-multiplier parameterization introduced in Sec.~\ref{sec:2}, while both its numerical realization and the SR component are adapted to the data representation.

\paragraph{Regular grids on a plane}
For regular grids on a plane, we directly use the base LSR-Net architecture and refer to this realization as \emph{p-LSR-Net}.
Since the data are sampled on a Cartesian mesh, the LR module admits a direct FFT implementation through its Fourier-multiplier form.
In the computations reported here, we retain the Fourier-multiplier/Gaussian-gridding pipeline for implementation consistency with the point-cloud settings; in the regular-grid case, this gridding step is optional rather than essential.
The SR component is implemented by standard local convolutions with compact stencils, which provide translation-equivariant local processing on the grid.

\paragraph{Point clouds on the sphere}
For spherical point clouds, we embed the LR--SR decomposition into the LSR-U-Net backbone and refer to this realization as \emph{s-LSR-Net}.
Because the sampled points on the sphere do not admit a direct FFT on the original discretization, the LR component is evaluated by the Gaussian-gridding procedure described in Sec.~\ref{sec:gaussiangridding} and Algorithm~\ref{alg1}.
The SR component is implemented by discrete--continuous (DISCO) convolution~\cite{ocampo2022disco}, which is designed to provide rotationally equivariant local processing on $\mathbb{S}^2$.
This property is important for spherical pattern dynamics, where the learned evolution should not depend on an arbitrary orientation of the coordinate frame.
Additional discussions of \emph{s-LSR-Net} are provided in~\ref{app:LSR-U-Net}, and a concise summary of DISCO is given in~\ref{app:sr-operators}.

\paragraph{Point clouds on general manifolds}
For point clouds sampled on general manifolds, we use a geometry-informed variant of the base LSR-Net, referred to as \emph{m-LSR-Net}.
In the LR branch, point-cloud features are first mapped to a latent reference grid by a geometry-informed GNO operator~\cite{li2023GINO}, which aggregates information from irregular samples using point locations and local neighborhood information.
The SOE-based LR module is then applied on this latent grid using the same Fourier-multiplier/Gaussian-gridding pipeline as above, and the resulting LR features are mapped back to the original point cloud.
In parallel, the SR branch is implemented by DeltaConv~\cite{wiersma2022deltaconv}, which operates directly on the manifold point cloud and defines intrinsic local interactions on discretized two-dimensional Riemannian manifolds embedded in $\mathbb{R}^3$.
A concise summary of DeltaConv and the geometry-informed GNO operator is also provided in~\ref{app:sr-operators}.


\subsection{Multi-step Training and Autoregressive Inference}

The problems considered here require stable long-time prediction. 
We therefore train the learned evolution operator in a multi-step setting and evaluate it through autoregressive inference. 
Starting from a state at time $t$, the model predicts the state at $t+T$; this prediction is then fed back as input to obtain the state at $t+2T$, and the procedure is repeated recursively.

We consider two multi-step training strategies. 
In autoregressive (AR) training~\cite{khurjekar2025enhanced}, the model is unrolled on its own intermediate predictions, so that the optimization directly accounts for error accumulation during inference. 
In teacher forcing (TF)~\cite{williams1989learning,bengio2015scheduled,lamb2016professor}, each one-step prediction is conditioned on the corresponding ground-truth state from the previous step. 
These strategies differ only during training; all reported test results are obtained by autoregressive inference.

For the planar phase-field problems on regular grids, including the Allen--Cahn and Cahn--Hilliard equations, we use AR training only. 
Preliminary experiments showed little difference between AR and TF in these cases, while AR already provided stable long-term predictions. 
To reduce variance, each planar model is trained independently with five random initializations and the predictions are averaged at each inference step.
For point clouds on the sphere and on general manifolds, both AR and TF results are reported when relevant. 
In these point-cloud settings, the preferred strategy depends on the dynamics and data representation. 
For single-field problems such as Allen--Cahn, AR and TF often perform similarly, while for coupled or more nonlinear systems, TF can yield smaller long-horizon prediction errors.

In all experiments, the step size $T$ is chosen from the early stage of pattern formation. 
The learned operator is then applied recursively to assess whether it can reproduce the subsequent pattern evolution over long time intervals.

\subsection{Evaluation Metrics}
To evaluate predictive performance, we use the relative mean-squared error, abbreviated as RMSE in the tables and figures. 
For predicted fields $\hat{\phi}^{(k)}$ and reference fields $\phi^{(k)}$ over $N_{\text{test}}$ test samples, we define
\begin{equation}
    \mathrm{RMSE}
    =
    \frac{
        \sum_{k=1}^{N_{\text{test}}}
        \bigl\| \hat{\phi}^{(k)} - \phi^{(k)} \bigr\|_2^2
    }{
        \sum_{k=1}^{N_{\text{test}}}
        \bigl\| \phi^{(k)} \bigr\|_2^2
    }.
\end{equation}
This relative normalization enables consistent comparison across systems with different physical scales. 
For fields sampled on regular grids and on point clouds over general manifolds, this discrete metric is used directly.

For fields defined on the sphere, we further introduce geometric weighting to account for the nonuniform surface area represented by the latitude--longitude discretization. 
Let $(\theta_i,\varphi_j)$ denote the spherical grid points, and let $q_i$ be the quadrature weights in the latitudinal direction. 
Under Gauss--Legendre quadrature, these weights are given by
\begin{equation}
    q_i
    =
    \frac{2}{(1-x_i^2)\,[P_N'(x_i)]^2},
    \qquad
    P_N(x_i)=0,
    \qquad
    \theta_i=\arccos x_i,
\end{equation}
where $x_i$ are the roots of the $N$th-order Legendre polynomial $P_N(x)$. 
The corresponding weight at each spherical grid point is $
    w_{ij} = q_i \,\Delta\varphi$, where $\Delta\varphi = 2\pi/N_\varphi$ is the longitudinal grid spacing. 
The weighted spherical RMSE is then defined by
\begin{equation}
    \mathrm{RMSE}_{\mathrm{sphere}}
    =
    \frac{
        \sum_{k=1}^{N_{\text{test}}}
        \sum_{i=1}^{N_\theta}
        \sum_{j=1}^{N_\varphi}
        w_{ij}\,
        \left|\hat{\phi}^{(k)}(\theta_i,\varphi_j)-\phi^{(k)}(\theta_i,\varphi_j)\right|^2
    }{
        \sum_{k=1}^{N_{\text{test}}}
        \sum_{i=1}^{N_\theta}
        \sum_{j=1}^{N_\varphi}
        w_{ij}\,
        \left|\phi^{(k)}(\theta_i,\varphi_j)\right|^2
    }.
\end{equation}
This weighting ensures that the discrete sum approximates the surface integral on $\mathbb{S}^2$ and avoids bias caused by nonuniform latitudinal sampling. 
This weighted metric is used throughout the spherical experiments.

\section{Numerical Results}\label{sec:results}

In this section, we evaluate LSR-Net on the three data representations considered in this work:
(i) regular grids on the Euclidean plane,
(ii) point clouds on the sphere, and
(iii) point clouds on general manifolds, represented here by a blob-shaped surface.
We also examine geometric consistency under rigid transformations for the spherical and general-manifold cases.
Details of the data preparation are provided in~\ref{app:Data Preparation}, and all model hyperparameter settings are summarized in~\ref{app:Model parameter settings}.

\subsection{Euclidean 2D Plane}

We first consider two representative phase-field models on regular Cartesian grids: the Allen--Cahn equation~\cite{AllenCahn1979} and the Cahn--Hilliard equation~\cite{CahnHilliard1958}. 
Phase-field models describe pattern-forming dynamics through a continuous order parameter $\phi(\mathbf{x},t)$ whose evolution is driven by gradients of a nonconvex free-energy landscape~\cite{cross1993pattern,steinbach2009phase}.
The Allen--Cahn equation takes the form
\begin{equation}
\frac{\partial \phi}{\partial t} = \epsilon^2 \nabla^2 \phi + (\phi - \phi^3),
\label{eq:allen_cahn}
\end{equation}
which is a reaction--diffusion equation commonly used to model interface motion and phase transitions. 
The Cahn--Hilliard equation reads
\begin{equation}
\frac{\partial \phi}{\partial t}
= M \nabla^2 \left[ -\kappa \nabla^2 \phi + 2W(\phi - 3\phi^2 + 2\phi^3) \right].
\label{eq:ch_case1}
\end{equation}

\begin{figure*}[htb!] 
    \centering
    \renewcommand{\arraystretch}{1.3}
    \begin{tabular}{c}
        \begin{tabular}{c @{\hspace{8mm}} c}  
            \begin{minipage}{0.45\textwidth}
                \centering
                  \includegraphics[width=\linewidth,keepaspectratio]{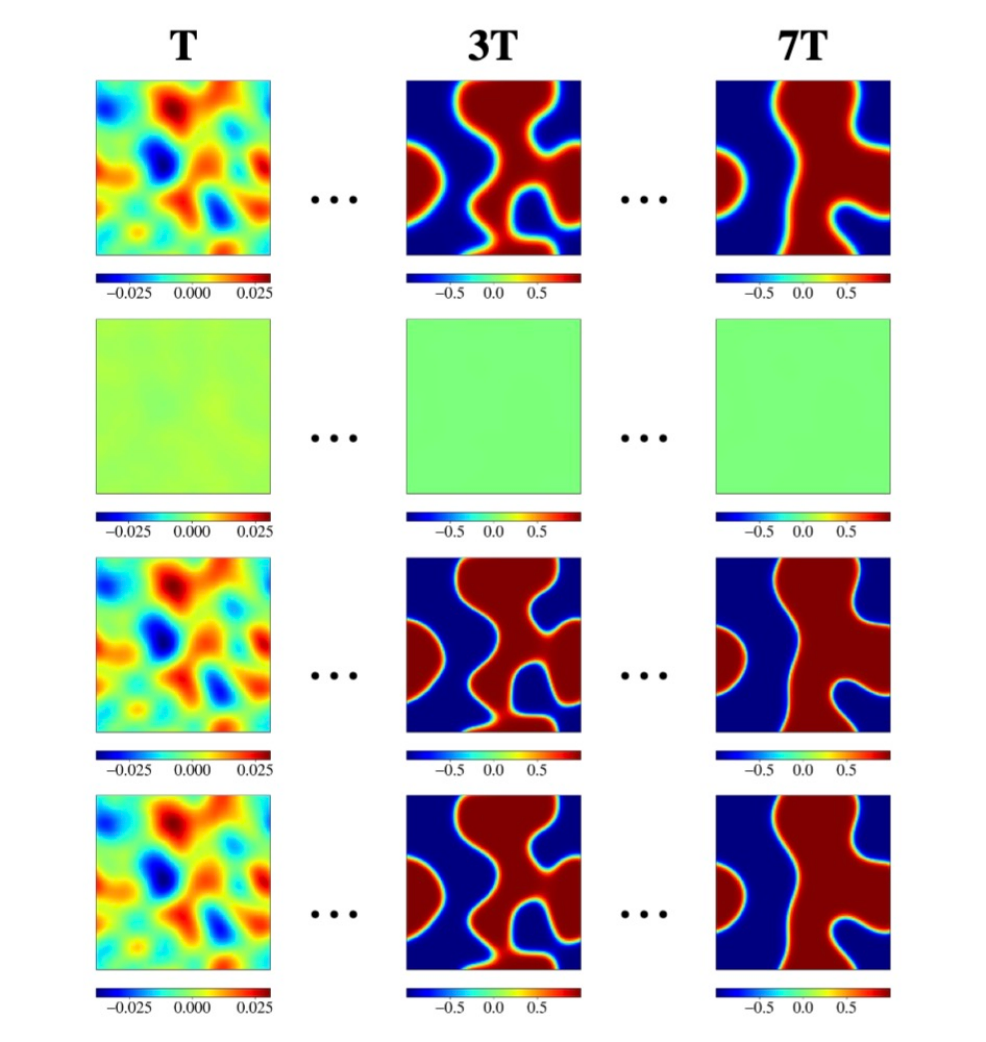}
                \textbf{(a) Allen--Cahn: Long-horizon predictions}
            \end{minipage}
            &
            \begin{minipage}{0.45\textwidth}
                \centering
                \includegraphics[width=\linewidth,keepaspectratio]{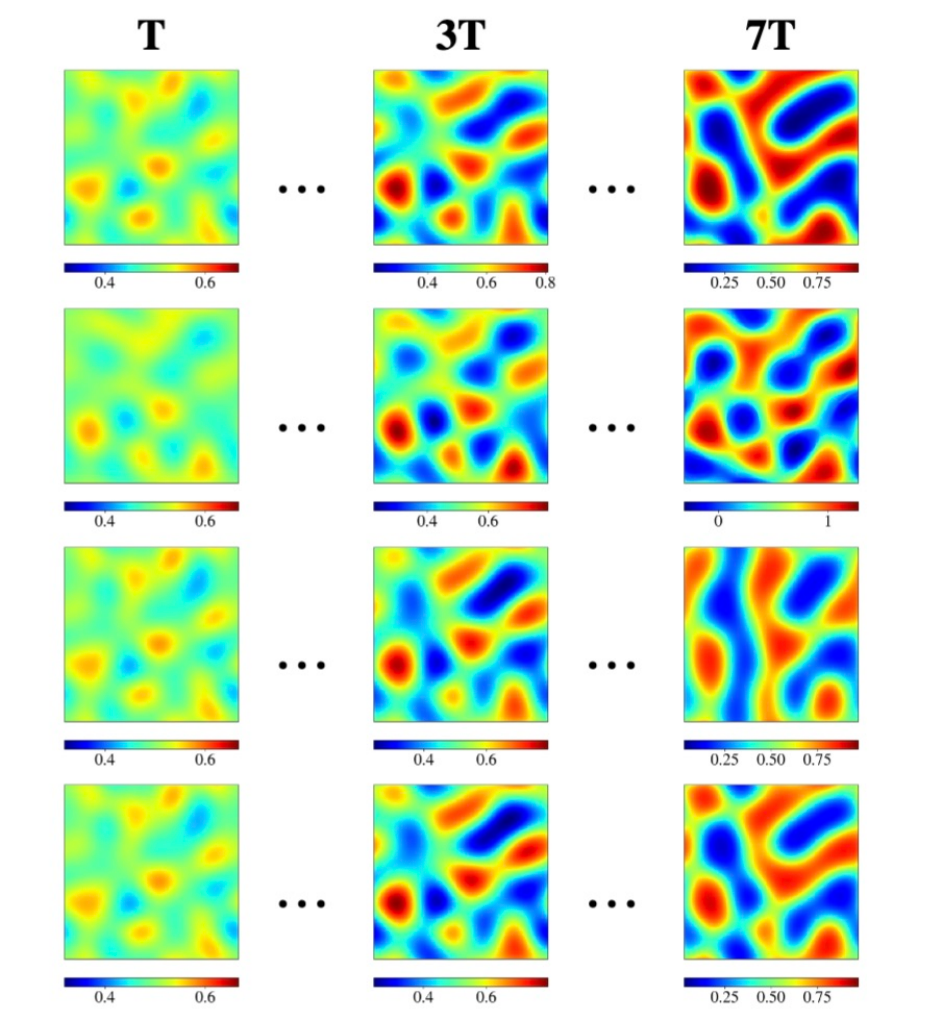}
                \textbf{(b) Cahn--Hilliard: Long-horizon predictions}
            \end{minipage}
        \end{tabular} \\[4mm]  

        \begin{tabular}{c @{\hspace{8mm}} c}
            \begin{minipage}{0.45\textwidth}
                \centering
                 \includegraphics[width=\linewidth,keepaspectratio]{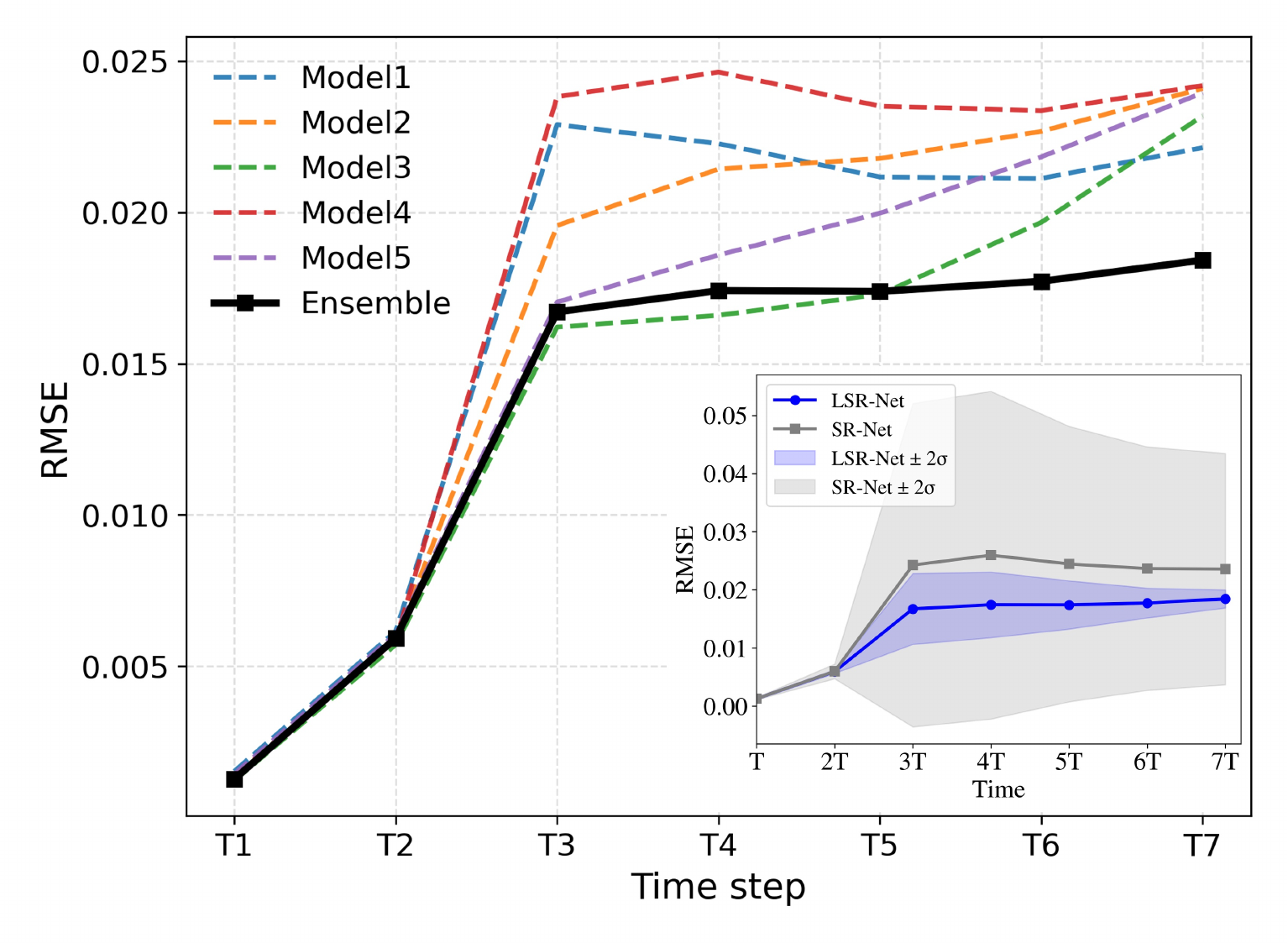}
                \textbf{(c) Allen--Cahn: Ensemble \& RMSE}
            \end{minipage}
            &
            \begin{minipage}{0.45\textwidth}
                \centering
                 \includegraphics[width=\linewidth,keepaspectratio]{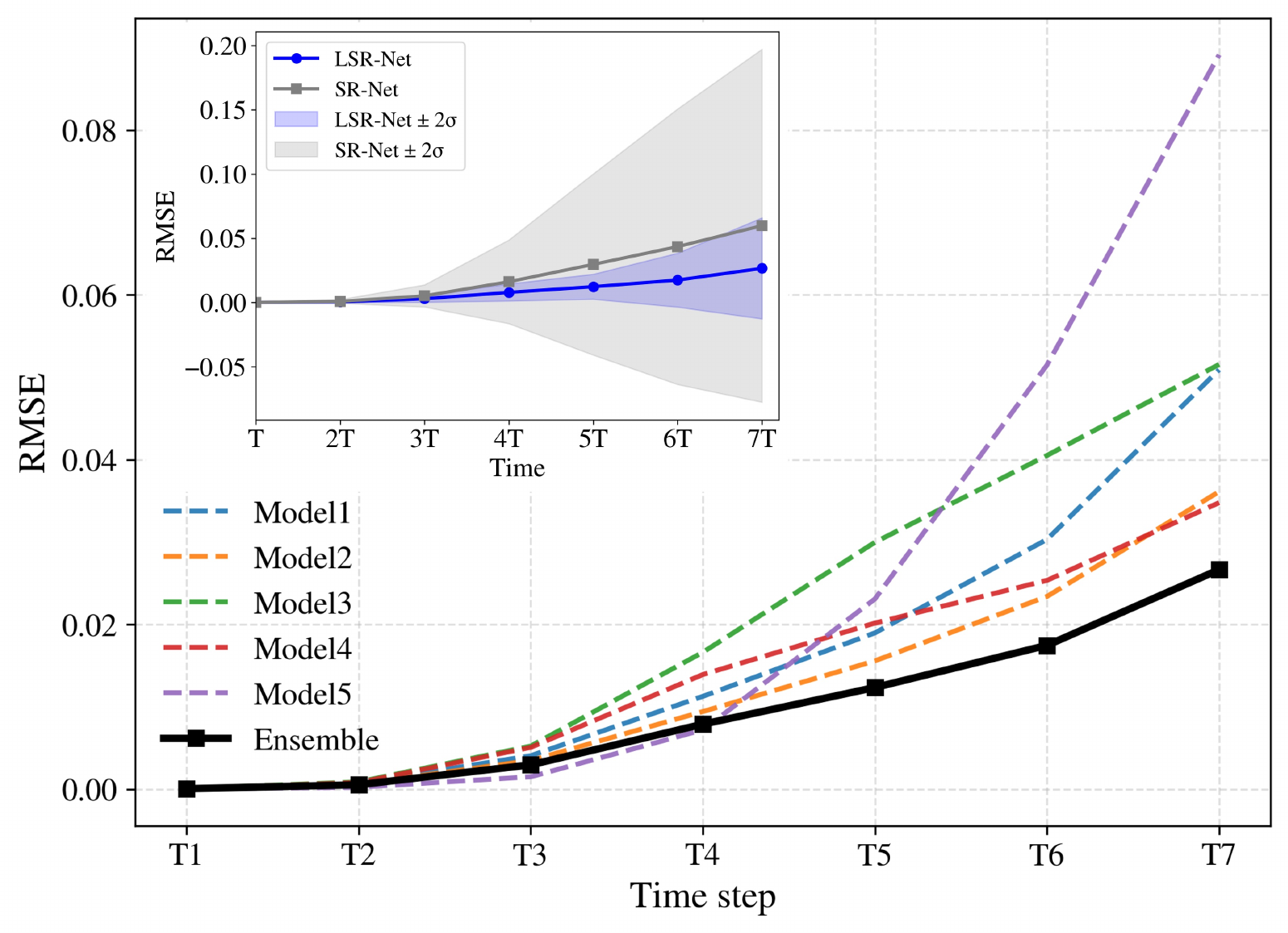}
                \textbf{(d) Cahn--Hilliard: Ensemble \& RMSE}
            \end{minipage}
        \end{tabular}
    \end{tabular}

    \caption{
        Long-horizon prediction results on regular grids.
        Top panel: representative autoregressive predictions for (a) Allen--Cahn and (b) Cahn--Hilliard at times $T$, $3T$, and $7T$.
        In each panel, the first row shows the reference solution, followed by predictions from FNO, SR-Net, and the p-LSR-Net.
        Bottom panel: ensemble predictions and RMSE statistics for (c) Allen--Cahn and (d) Cahn--Hilliard.
        Insets show the RMSE evolution during autoregressive inference; solid curves denote ensemble means and shaded bands indicate two standard deviations across five independent runs.
        SR-Net corresponds to the ablated model in which the long-range module is removed.
    }
    \label{fig:ac_ch_combined_full}
\end{figure*}

Figs.~\ref{fig:ac_ch_combined_full}(a,b) show representative autoregressive predictions at times $T$, $3T$, and $7T$. 
The first row in each panel gives the reference solution, followed by predictions from FNO, SR-Net, and p-LSR-Net. 
Here SR-Net denotes the ablated model obtained by removing the LR module from p-LSR-Net. 
For each method, predictions are averaged over five independently trained models to reduce variance due to random initialization and stochastic optimization.

The ensemble behavior and RMSE evolution are shown in Figs.~\ref{fig:ac_ch_combined_full}(c,d), where solid curves denote ensemble-averaged errors and shaded regions indicate two standard deviations ($2\sigma$) across the five runs. 
For Allen--Cahn, FNO deteriorates rapidly under autoregressive inference, whereas both SR-Net and p-LSR-Net remain stable. 
For Cahn--Hilliard, the difference is more pronounced: p-LSR-Net better preserves both large-scale morphology and fine-scale structure over long horizons. 
The corresponding RMSE curves show slower error growth and smaller run-to-run variability for p-LSR-Net than for the baseline models.

\begin{table}[htbp]
    \centering
    \caption{RMSE comparison of FNO, SR-Net, and p-LSR-Net(AR) for the Allen--Cahn ($T=4$) and Cahn--Hilliard ($T=7$) systems at different autoregressive prediction horizons.}
    \begin{tabular}{c|c|ccccccc}
        \hline
        System & Method & $T$ & $2T$ & $3T$ & $4T$ & $5T$ & $6T$ & $7T$ \\
        \hline
        \multirow{3}{*}{Allen--Cahn}
        & FNO     
        & 0.96959 & 0.99965 & 0.99998 & 1.00002 & 1.00003 & 1.00004 & 1.00005 \\
        & SR-Net      
        & \textbf{0.00126} & 0.00597 & 0.02424 & 0.02596 & 0.02443 & 0.02364 & 0.02357 \\
        & p-LSR-Net    
        & 0.00127 & \textbf{0.00593} & \textbf{0.01671} & \textbf{0.01741} & \textbf{0.01739} & \textbf{0.01772} & \textbf{0.01842} \\
        \hline
        \multirow{4}{*}{Cahn--Hilliard}
        & FNO     
        & 0.00057 & 0.01679 & 0.05280 & 0.12015 & 0.18293 & 0.22297 & 0.24926 \\
        & SR-Net      
        & 0.00009 & 0.00087 & 0.00520 & 0.01618 & 0.02968 & 0.04350 & 0.05982 \\
        & p-LSR-Net   
        & \textbf{0.00006} & \textbf{0.00054} & \textbf{0.00296} & \textbf{0.00790} & \textbf{0.01235} & \textbf{0.01747} & \textbf{0.02665} \\
        \hline
    \end{tabular}
    \label{tab:rmse_comparison}
\end{table}

Table~\ref{tab:rmse_comparison} summarizes the RMSE values for FNO, SR-Net, and p-LSR-Net. 
p-LSR-Net attains the smallest error at most prediction horizons, and the advantage becomes more pronounced as the horizon increases. 
For Cahn--Hilliard, FNO exhibits strong error accumulation under autoregressive inference, whereas p-LSR-Net maintains substantially smaller errors throughout the rollout. 
For Allen--Cahn, FNO becomes inaccurate after only a few steps, while SR-Net and p-LSR-Net remain stable, with p-LSR-Net achieving the smallest long-horizon RMSE for predictions above $2T$. 
These results indicate that adding the LR module improves stable long-time prediction for dissipative pattern-forming systems.

\subsection{Point Clouds on a Sphere}

We next evaluate the spherical realization of LSR-Net, i.e., the s-LSR-Net, for point cloud data sampled from PDEs defined on $\mathbb{S}^2$. 
Unless otherwise stated, test trajectories are evaluated by autoregressive inference, and errors are reported using the weighted spherical RMSE defined in Sec.~\ref{sec:3}. 
We compare against the spherical Fourier neural operator (SFNO)~\cite{bonev2023spherical} and report both AR- and TF-trained s-LSR-Net variants when applicable.

\begin{figure*}[htb!]
    \centering
    \renewcommand{\arraystretch}{0.8}
    \setlength{\tabcolsep}{1pt}
    \scriptsize

     \includegraphics[width=0.8\textwidth]{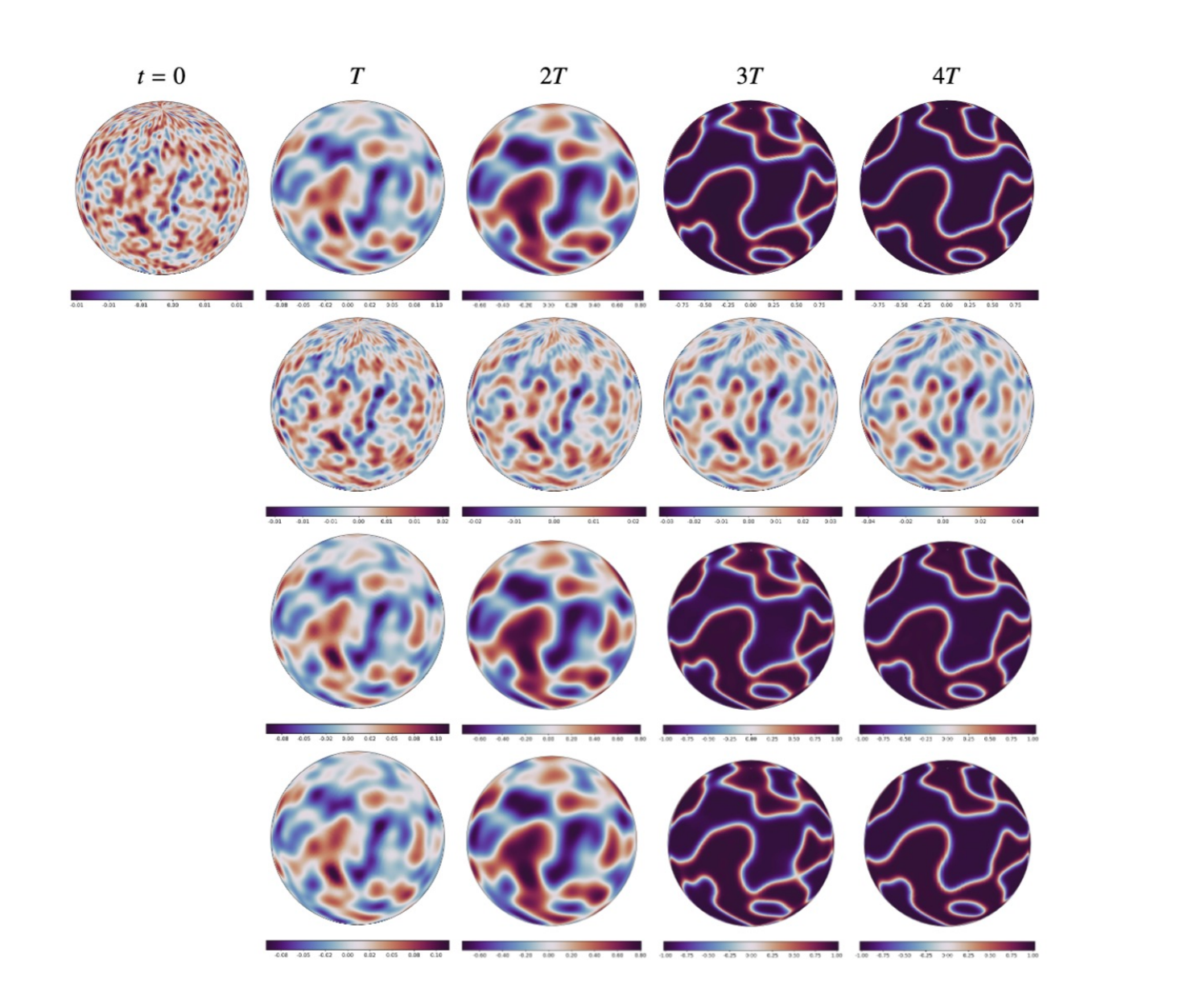}

    \caption{
        Allen--Cahn dynamics on the sphere.
        The leftmost column shows the input at \(t=0\), followed by the reference and predicted solutions at times $T$, $2T$, $3T$, and $4T$($T=3$).
        The four rows correspond to the reference solution, SFNO, s-LSR-Net(AR), and s-LSR-Net(TF) predictions, respectively.
    }
    \label{fig:ac_sphere_infer}
\end{figure*}

We first consider the Allen--Cahn equation on the sphere, using the same governing dynamics as in Eq.~\eqref{eq:allen_cahn}, where the Laplacian is replaced by the Laplace--Beltrami operator~\cite{epstein2010debye,o2018second}. 
For this single-field problem, we examine both AR and TF training strategies.
Fig.~\ref{fig:ac_sphere_infer} presents representative predictions at times $T$, $2T$, $3T$, and $4T$ with $T=3$. 
The first row shows the input and reference solutions, followed by predictions from SFNO, s-LSR-Net(AR), and s-LSR-Net(TF). 
The s-LSR-Net results with both AR and TF training strategies accurately capture the phase-field evolution, preserving sharp interfaces and global structures over long horizons. 
In contrast, SFNO rapidly deviates from the reference trajectory.

Table~\ref{tab:ac_sphere_rmse_comparison} reports the corresponding weighted spherical RMSE values. 
SFNO exhibits rapid error growth, with RMSE values already close to $O(1)$ by $2T$. 
In contrast, both s-LSR-Net(AR) and s-LSR-Net(TF) remain stable and accurate across all reported horizons, with errors nearly three orders of magnitude smaller than those of SFNO. 
The two training strategies perform similarly on this problem, with the AR variant showing slightly better long-horizon stability at $4T$. 
This behavior is consistent with the discussion in Sec.~\ref{sec:3}: for this single-field spherical problem, either AR or TF training can be effective.

\begin{table}[htbp]
    \centering
    \caption{Weighted spherical RMSE of SFNO, s-LSR-Net(AR), and s-LSR-Net(TF) for Allen--Cahn on the sphere ($T=3$).}
    \begin{tabular}{c|cccc}
        \hline
        Method & $T$ & $2T$ & $3T$ & $4T$ \\
        \hline
        SFNO             & 0.7839 & 0.9787 & 0.9909 & 0.9897 \\
       s-LSR-Net(AR)     & 0.0005 & \textbf{0.0004} & \textbf{0.0006} & \textbf{0.0010} \\
        s-LSR-Net(TF)     & \textbf{0.0004} & 0.0005 & 0.0010 & 0.0017 \\
        \hline
    \end{tabular}
    \label{tab:ac_sphere_rmse_comparison}
\end{table}

We next consider the classical Schnakenberg reaction--diffusion system~\cite{schnakenberg1979simple} defined on a spherical surface $\Gamma$,
\begin{equation}
\begin{cases}
\displaystyle \frac{\partial u}{\partial t}
= D_u \Delta_\Gamma u + a - u + u^2 v, \\[0.6em]
\displaystyle \frac{\partial v}{\partial t}
= D_v \Delta_\Gamma v + b - u^2 v,
\end{cases}
\label{eq:schnakenberg}
\end{equation}
where $u$ and $v$ denote the concentrations of two interacting species, $D_u$ and $D_v$ are diffusion coefficients, $a$ and $b$ are reaction parameters, and $\Delta_\Gamma$ is the Laplace--Beltrami operator on the surface. 
This problem is more challenging than the spherical Allen--Cahn case since it is both nonlinear and involves coupled multi-species dynamics, making it a useful test to distinguish the AR/TF training strategies discussed in Sec.~\ref{sec:3}.

\begin{figure*}[htb!]
    \centering
    \renewcommand{\arraystretch}{0.8}
    \setlength{\tabcolsep}{1pt}
    \scriptsize
       \includegraphics[width=0.9\textwidth]{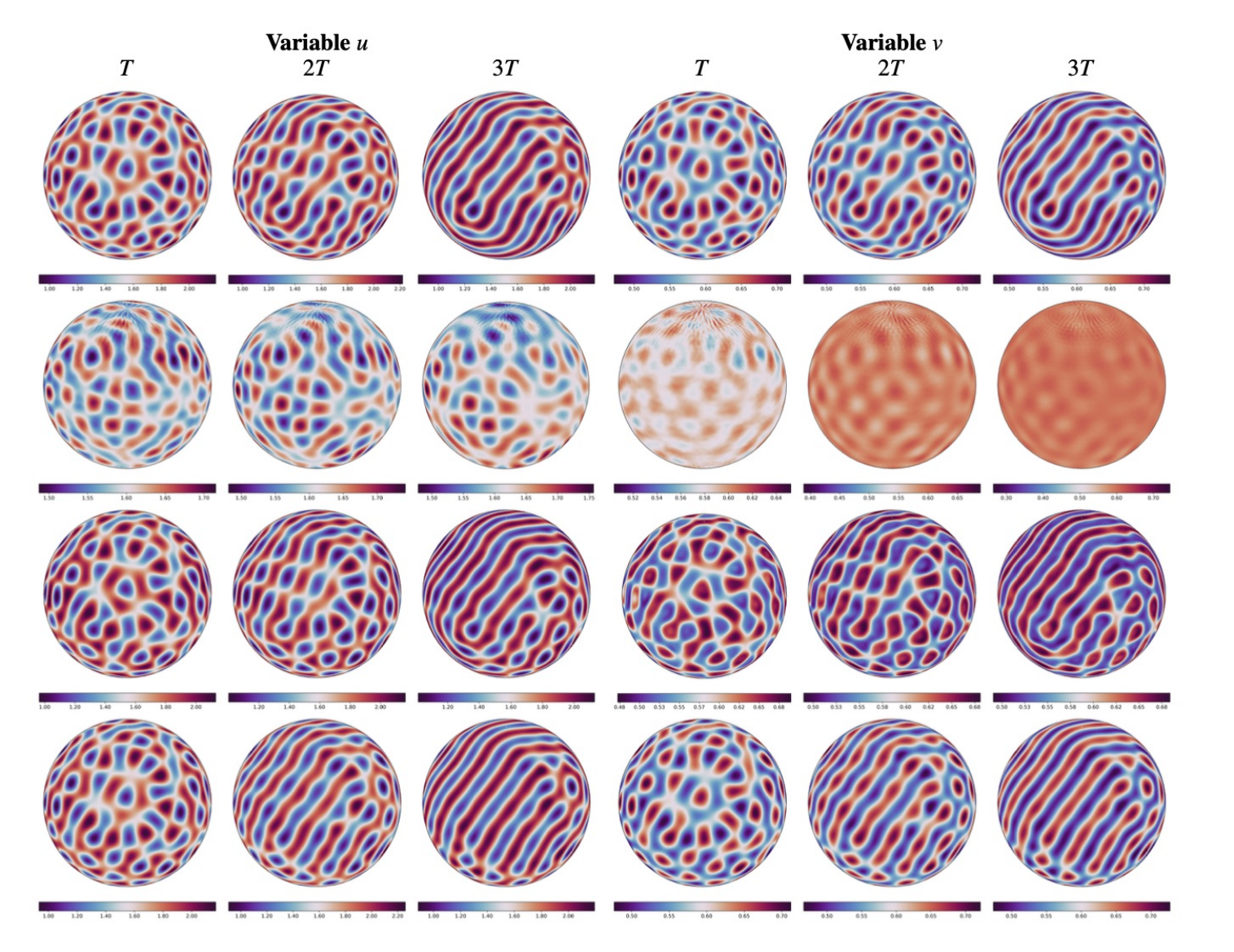}
    \caption{
        Predicted stripe patterns of the Schnakenberg system for variables $u$ (left) and $v$ (right) at times $T$, $2T$, and $3T$ ($T=260$).
        The four rows correspond to the reference solution, SFNO, s-LSR-Net(AR), and s-LSR-Net(TF) predictions, respectively.
    }
    \label{fig:schnaken_infer_uv}   
\end{figure*}

Fig.~\ref{fig:schnaken_infer_uv} reports predictions at $T$, $2T$, and $3T$ for both species $u$ and $v$. 
The four rows correspond to the reference solution, SFNO, s-LSR-Net(AR), and s-LSR-Net(TF) predictions at $T$, $2T$ and $3T$ ($T=260$), respectively. 
SFNO fails to reproduce the stripe-like patterns consistently, especially for the $v$ field, where the predicted patterns quickly lose fidelity. 
In contrast, s-LSR-Net with either AR or TF training captures the temporal evolution more accurately while preserving both large-scale structures and fine spatial details.
Here the TF-trained model is slightly more accurate than the AR-trained model, although the AR result remains stable as well.

Table~\ref{tab:schnakenberg_rmse} confirms the trends observed in Fig.~\ref{fig:schnaken_infer_uv} quantitatively. 
SFNO exhibits large errors starting from the first prediction step, reflecting the difficulty of learning these coupled spherical dynamics with this baseline model. 
The s-LSR-Net results with both AR and TF training strategies reduce the error by orders of magnitude compared to SFNO, and the TF-trained model consistently attains the lowest RMSE over the reported horizons. 
This example validates the discussions in Sec.~\ref{sec:3}, namely, the TF training strategy is found to provide an advantage for multi-species coupled pattern dynamics.

\begin{table}[htbp]
    \centering
    \caption{Weighted spherical RMSE of s-LSR-Net and SFNO on the Schnakenberg system ($T=260$).}
    \begin{tabular}{c|ccc}
        \hline
        Method            & $T$      & $2T$      & $3T$      \\
        \hline
        SFNO              & 0.0229        & 0.0295        & 0.0328        \\
        s-LSR-Net(AR)      & 0.0017        & 0.0043        & 0.0076        \\
        s-LSR-Net(TF)      & \textbf{0.0004} & \textbf{0.0015} & \textbf{0.0039} \\
        \hline
    \end{tabular}
    \label{tab:schnakenberg_rmse}
\end{table}

We also consider the spherical Turing system; its full formulation and additional results are reported in~\ref{app:Turing}. 
As in the Schnakenberg case, s-LSR-Net results with both AR and TF training strategies substantially outperform SFNO; TF yields slightly smaller short-horizon errors, while AR remains stable at longer horizons as well.

\subsection{Point Clouds on General Manifolds}

We finally evaluate the general-manifold realization of LSR-Net (m-LSR-Net) on point clouds sampled from a blob-shaped surface. 
The test problem is the Allen--Cahn equation posed on a smooth, closed surface 
$\Gamma \subset \mathbb{R}^3$:
\begin{equation}
    \partial_t u = \delta \Delta_{\Gamma} u + u - u^3,
    \label{eq:ac_surface}
\end{equation}
where $u=u(x,t)$ denotes the order parameter, $\delta>0$ is the diffusion coefficient, and $\Delta_\Gamma$ is the Laplace--Beltrami operator. 
Because the phase-field dynamics is constrained to the surface geometry, local curvature effects can influence interface motion and pattern formation.
We compare m-LSR-Net with GINO~\cite{li2023GINO} and with an SR-only baseline, denoted SR-Net(DeltaConv), in which DeltaConv~\cite{wiersma2022deltaconv} provides the short-range operator without the LR branch. Note that detailed data generation and model settings are given in~\ref{app:Data Preparation} and~\ref{app:Model parameter settings}.

\begin{figure*}[htb!] 
    \centering
    \renewcommand{\arraystretch}{1.0}
    \setlength{\tabcolsep}{2pt}
    \scriptsize
   \includegraphics[width=0.8\textwidth]{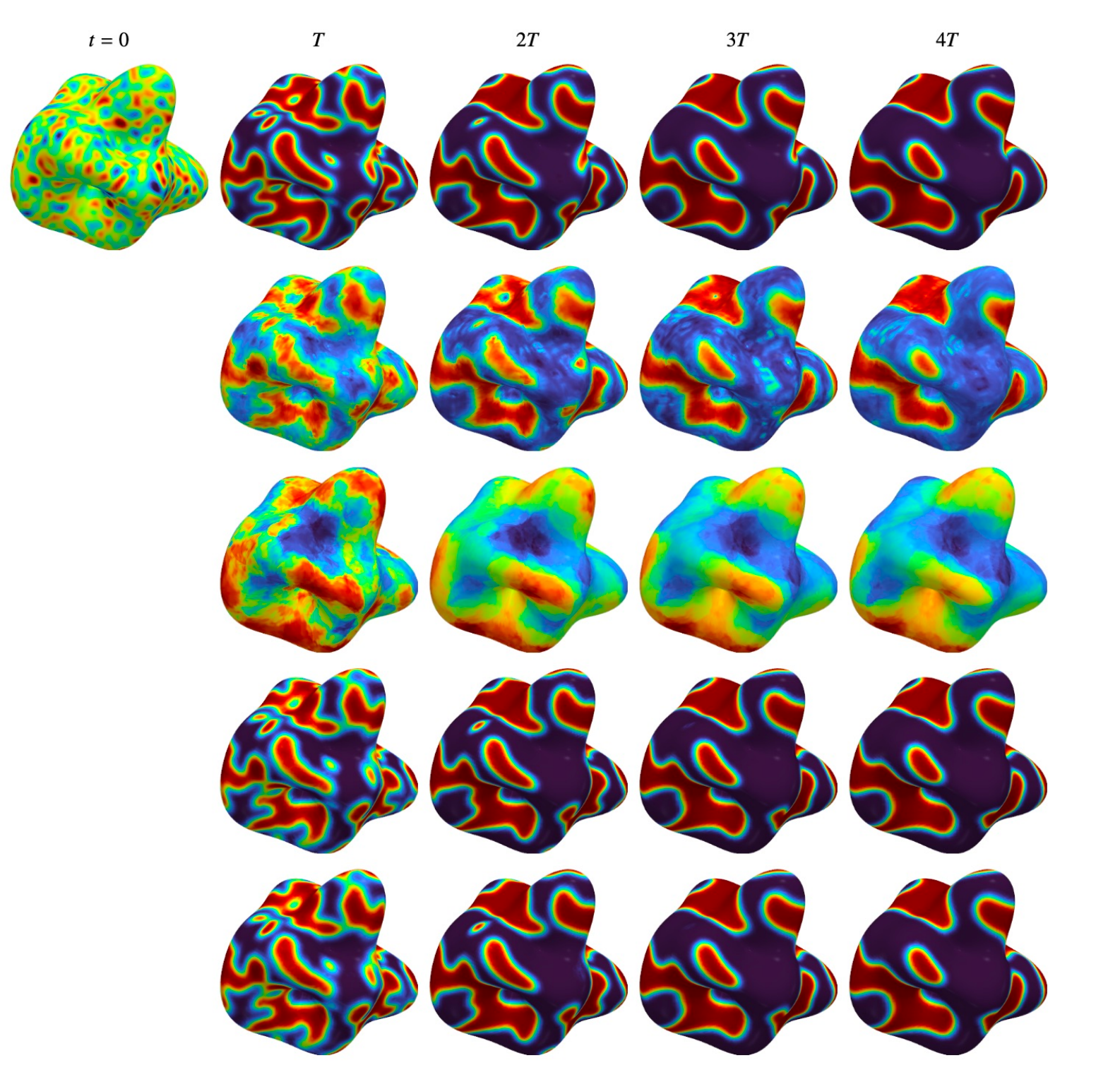}
    \caption{
        Point-cloud predictions of the Allen--Cahn equation on a blob-shaped manifold.
        The figure shows the input at $t=0$, followed by the reference solution and model predictions at times $T$, $2T$, $3T$, and $4T$ ($T=5$).
        The first row presents the input and reference evolution, and the remaining rows show the predictions of SR-Net(DeltaConv), GINO(TF), m-LSR-Net(AR), and m-LSR-Net(TF), respectively.
    }
    \label{fig:ac_point_infer}
\end{figure*}

Fig.~\ref{fig:ac_point_infer} compares autoregressive predictions on the blob-shaped manifold.
Compared with GINO and SR-Net(DeltaConv), m-LSR-Net better captures the phase morphology and interface structure throughout the rollout. 
The difference becomes more pronounced at later horizons: both GINO and SR-Net(DeltaConv) exhibit substantial error accumulations,  while m-LSR-Net predictions remain close to the reference evolution.
\begin{table}[htbp!]
    \centering
    \caption{RMSE of SR-Net(DeltaConv), GINO, and m-LSR-Net on Allen--Cahn over the blob-shaped manifold ($T=5$).}
    \begin{tabular}{c|cccc}
        \hline
        Method          & $T$       & $2T$      & $3T$      & $4T$      \\
        \hline
        SR-Net(DeltaConv)       & 0.25140   & 0.19890   & 0.25763   & 0.26526   \\
        GINO(TF)        & 0.70043   & 0.82703   & 0.89677   & 0.93346   \\
        m-LSR-Net(AR)        & 0.02551   & 0.04155   & 0.06279   & 0.08045   \\
       m-LSR-Net(TF)        & \textbf{0.02238} & \textbf{0.03440} & \textbf{0.05233} & \textbf{0.06772} \\
        \hline
    \end{tabular}
    \label{tab:glsr_delta_auto_mse}
\end{table}
Table~\ref{tab:glsr_delta_auto_mse} reports the corresponding multi-step prediction errors and confirms the observed trends in Fig.~\ref{fig:ac_point_infer}. 
GINO(TF) has the largest errors, which continue to increase as the prediction horizon extends, indicating difficulty in maintaining long-time dynamics on this geometry. 
SR-Net(DeltaConv) reduces the error relative to GINO, but its accuracy remains limited without the LR branch. 
Both m-LSR-Net variants achieve substantially smaller errors across all reported horizons, with the TF-trained model giving the lowest RMSE and the AR-trained model also showing stable error growth during autoregressive inference.

\subsection{Geometric Consistency Under Rigid Transformations}
\label{sec:geom-consistency}

Predictive accuracy alone does not determine whether the learned dynamics are consistent with the underlying geometry. 
We therefore test whether the models produce compatible predictions when the input point clouds are transformed by rigid motions. 
For spherical data, we examine rotations of the input field; for general-manifold point clouds, we examine both rotations and reflections.

\subsubsection{Spherical Rotation Consistency of s-LSR-Net}
\label{sec:sphere-rotation-consistency}

We first assess spherical rotation consistency for s-LSR-Net. 
The test compares two inference paths: predicting directly from the original input, and predicting from a rotated input followed by rotating the output back to the original frame. 
The discrepancy between the two predictions measures the equivariance error introduced by the full numerical pipeline, including interpolation and autoregressive rollout.

Given a scalar field $u(\theta,\phi)$ on the unit sphere, we write the spherical-to-Cartesian map as
\begin{equation}
\mathbf{x}(\theta,\phi)=
\begin{bmatrix}
\sin\theta \cos\phi\\
\sin\theta \sin\phi\\
\cos\theta
\end{bmatrix}.
\end{equation}
A rigid rotation by angle $\alpha$ about a prescribed axis is then written as
\begin{equation}
\mathbf{x}' = \mathbf{R}_\alpha \mathbf{x},
\end{equation}
with
\begin{equation}
\mathbf{R}_x(\alpha)=\begin{bmatrix}1&0&0\\0&\cos\alpha&\sin\alpha\\0&-\sin\alpha&\cos\alpha\end{bmatrix},\ 
\mathbf{R}_y(\alpha)=\begin{bmatrix}\cos\alpha&0&-\sin\alpha\\0&1&0\\\sin\alpha&0&\cos\alpha\end{bmatrix},\ 
\mathbf{R}_z(\alpha)=\begin{bmatrix}\cos\alpha&\sin\alpha&0\\-\sin\alpha&\cos\alpha&0\\0&0&1\end{bmatrix}.
\end{equation}
The rotated coordinates are mapped back to spherical variables by
\begin{equation}
\theta' = \arccos(z'), \qquad 
\phi' = \mathrm{atan2}(y', x').
\end{equation}
After inference, the predicted fields are rotated back with the inverse transformation and compared with the model predictions from the original unrotated inputs.

\begin{figure*}[htb!] 
    \centering
    \renewcommand{\arraystretch}{1.0}
    \setlength{\tabcolsep}{2pt}
    \scriptsize
    \includegraphics[width=0.8\textwidth]{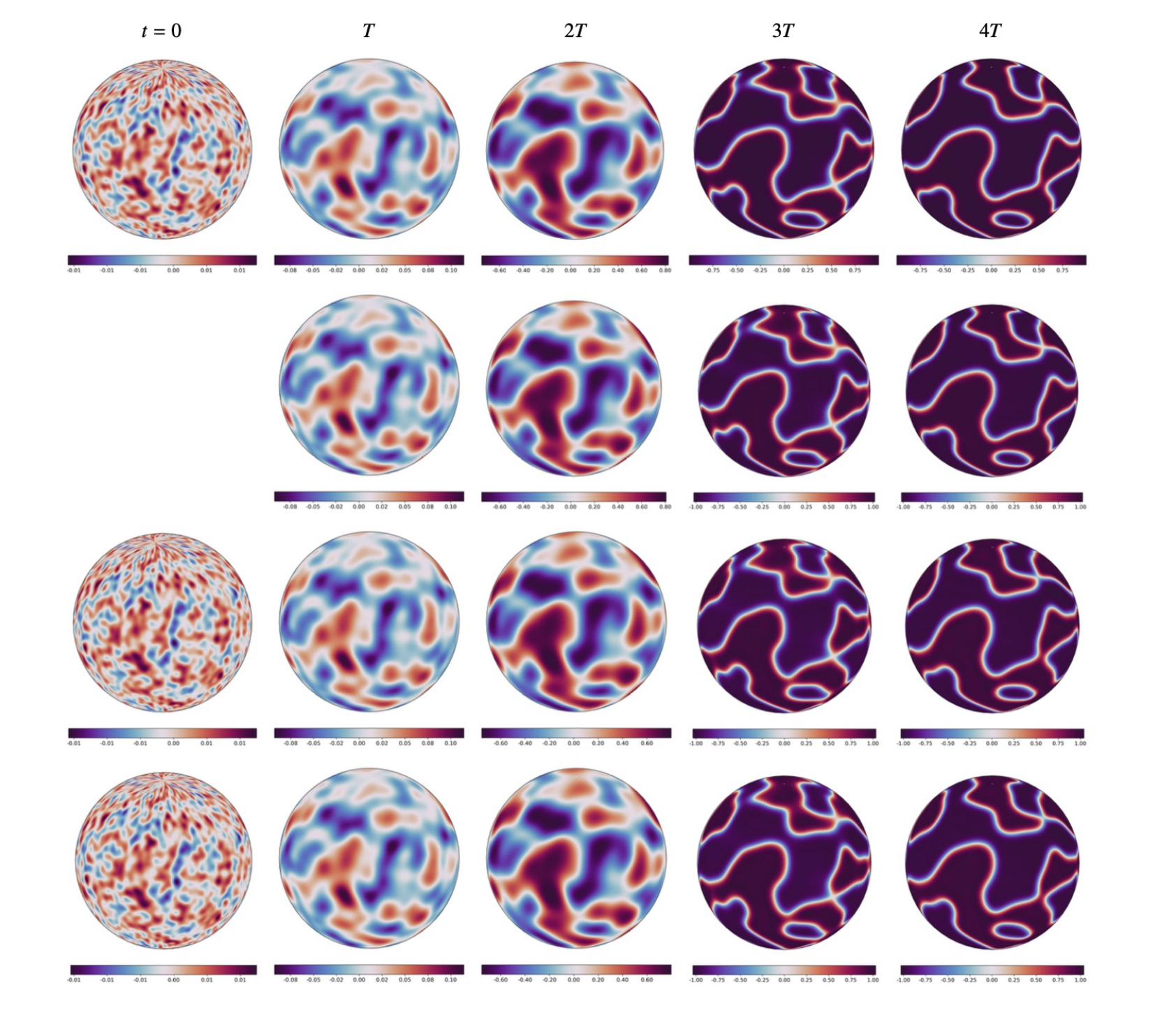}
	    \caption{
        Spherical rotation consistency of s-LSR-Net for the Allen--Cahn equation.
        The first row shows the input at $t=0$ and the reference solutions at times $T$, $2T$, $3T$, and $4T$ ($T=3$).
        The second row shows predictions from the original input, the third row shows predictions from inputs rotated by $5^\circ$ about the $x$-axis, and the fourth row shows the rotated-back predictions in the original frame.
        The close agreement between the second and fourth rows indicates that long-horizon predictions remain stable under rigid rotations.
	    }
	    \label{fig:ac_rotation_x_shpere_infer}
\end{figure*}

Fig.~\ref{fig:ac_rotation_x_shpere_infer} illustrates a representative example in which the spherical input is rotated about the $x$-axis. 
The original and rotated-back predictions remain closely aligned over the rollout, indicating that s-LSR-Net preserves the solution structure under rigid rotations.

Table~\ref{tab:ac_sphere_rotation_rmse} further quantifies this behavior. 
The ``Original'' rows report the prediction RMSE for unrotated inputs, while the rotated rows report the RMSE between the original predictions and the corresponding rotated-back predictions. 
For both $5^\circ$ and $15^\circ$ rotations, the consistency errors remain at the same small scale as the unrotated prediction errors, showing that the spherical implementation introduces only negligible rotational-invariance discrepancy.

\begin{table}[htbp]
    \centering
    \caption{Rotation-consistency results for s-LSR-Net(TF) on the spherical Allen--Cahn equation with $T=3$. The ``Original'' rows report prediction RMSE for unrotated inputs, while the rotated rows report the RMSE between the original predictions and the corresponding rotated-back predictions. Smaller rotated-row values indicate stronger consistency under rigid rotations.}
    \begin{tabular}{cc|cccc}
        \hline
        Rotation axis & Angle ($^\circ$) & $T$ & $2T$ & $3T$ & $4T$ \\
        \hline
        Original & 0 
        & $4.37\times10^{-4}$ 
        & $4.76\times10^{-4}$ 
        & $8.07\times10^{-4}$ 
        & $1.11\times10^{-3}$ \\
         \hline
        $x$ & 5 
        & $2.55\times10^{-4}$ 
        & $2.90\times10^{-4}$ 
        & $5.80\times10^{-4}$ 
        & $9.10\times10^{-4}$ \\

        $y$ & 5
        & $1.22\times10^{-4}$ 
        & $1.35\times10^{-4}$ 
        & $1.95\times10^{-4}$ 
        & $3.37\times10^{-4}$ \\

        $z$ & 5
        & $3.19\times10^{-4}$ 
        & $3.78\times10^{-4}$ 
        & $8.25\times10^{-4}$ 
        & $1.22\times10^{-3}$ \\
        \hline

         \hline
        Original & 0 
        & $4.63\times10^{-4}$ 
        & $6.71\times10^{-4}$ 
        & $1.24\times10^{-3}$ 
        & $1.94\times10^{-3}$ \\
         \hline
        $x$ & 15
        & $4.07\times10^{-4}$ 
        & $5.26\times10^{-4}$ 
        & $9.33\times10^{-4}$ 
        & $1.39\times10^{-3}$ \\

        $y$ & 15
        & $2.42\times10^{-4}$ 
        & $3.38\times10^{-4}$ 
        & $6.41\times10^{-4}$ 
        & $1.10\times10^{-3}$ \\

        $z$ & 15
        & $3.19\times10^{-4}$ 
        & $3.59\times10^{-4}$ 
        & $7.36\times10^{-4}$ 
        & $1.12\times10^{-3}$ \\
        \hline    
    \end{tabular}
    \label{tab:ac_sphere_rotation_rmse}
\end{table}

\subsubsection{Rotation and Reflection Consistency of m-LSR-Net}
\label{sec:manifold-consistency}

We next apply the same consistency test to m-LSR-Net on irregular point clouds sampled from a general manifold. 
For rotations, we use the same canonical-axis rotations introduced above and apply them directly to the point-cloud coordinates. 
We also examine the consistency of m-LSR-Net with respect to reflections. Given a point cloud
\(
\mathbf{X} = \{ \mathbf{x}_i \in \mathbb{R}^3 \}_{i=1}^N
\)
with an associated scalar field
\(
\mathbf{u} = \{ u_i \}_{i=1}^N,
\)
we reflect the coordinates with respect to the canonical coordinate planes.
Specifically, reflections with respect to the $xy$-, $xz$-, and $yz$-planes are defined by
\begin{equation}
\mathcal{M}_{xy}(x,y,z) = (x,y,-z), \quad
\mathcal{M}_{xz}(x,y,z) = (x,-y,z), \quad
\mathcal{M}_{yz}(x,y,z) = (-x,y,z).
\end{equation}
The reflection is applied to the point coordinates, while the scalar values are transported with their corresponding points:
\begin{equation}
\mathbf{X}' = \mathcal{M}(\mathbf{X}), \qquad
u'(\mathcal{M}(\mathbf{x}_i)) = u(\mathbf{x}_i),
\end{equation}
where $\mathcal{M}$ denotes the chosen reflection.

\begin{figure*}[htbp!] 
    \centering
    \renewcommand{\arraystretch}{1.0}
    \setlength{\tabcolsep}{2pt}
    \scriptsize
\includegraphics[width=1\textwidth]{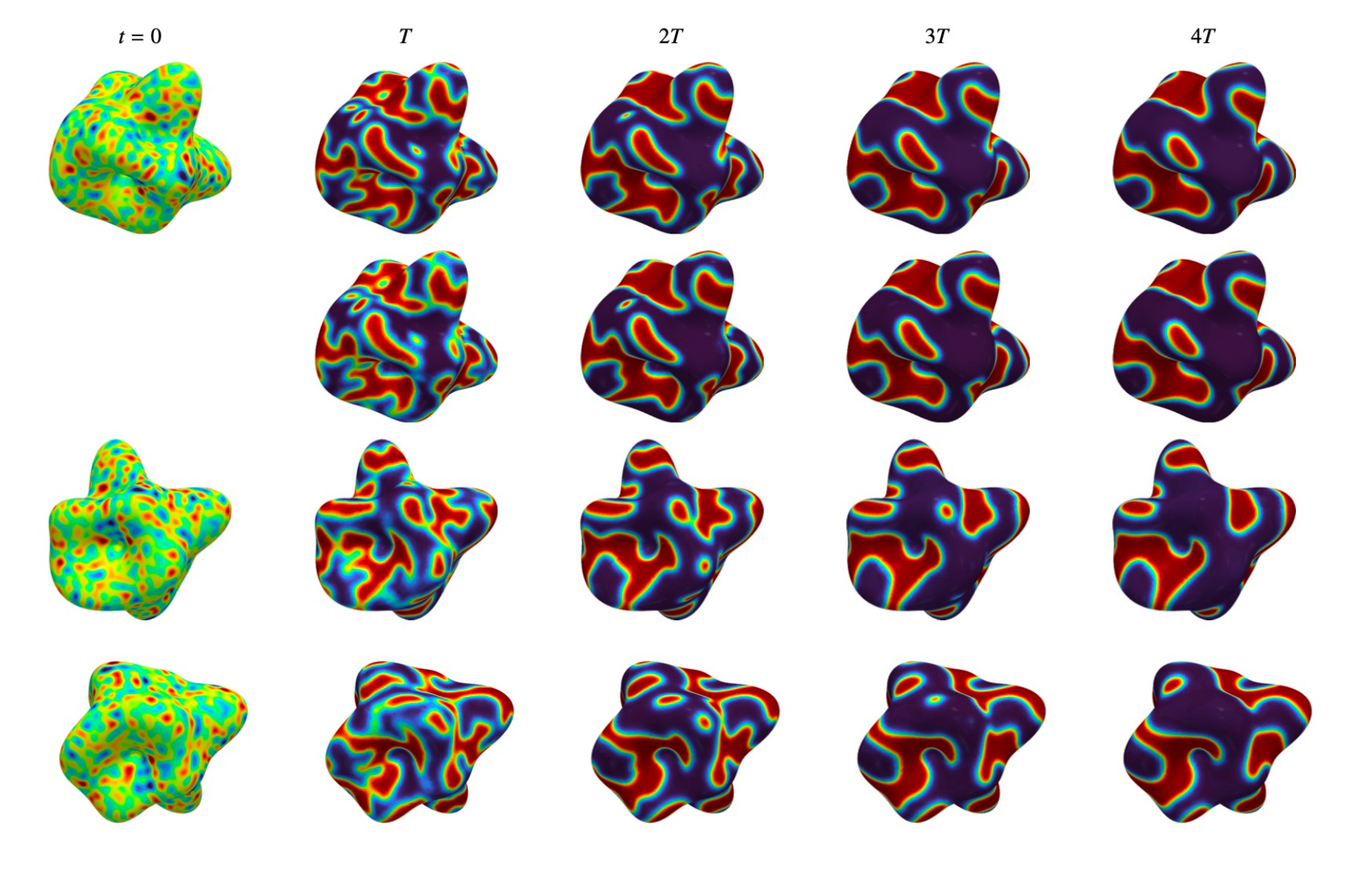}
	    
	    \caption{
        Rotation and reflection consistency of m-LSR-Net for the Allen--Cahn equation on irregular point clouds.
        The first row shows the input at $t=0$ and the reference solutions at times $T$, $2T$, $3T$, and $4T$ ($T=5$).
        The second row shows predictions from the original input, the third row shows predictions after a $60^\circ$ rotation about the $x$-axis, and the fourth row shows predictions after reflection with respect to the $xy$-plane.
        The preserved solution patterns across the transformed cases indicate that m-LSR-Net remains geometrically consistent under both rigid rotations and reflections.
        }
	    \label{fig:ac_isometrypoint_infer}
\end{figure*}

Fig.~\ref{fig:ac_isometrypoint_infer} shows representative rotation and reflection test results. 
The transformed predictions preserve the same phase morphology as the original rollout, despite the change in point-cloud coordinates.
Table~\ref{tab:ac_symmetry_rmse} reports the corresponding consistency errors. 
The ``Original'' row gives the prediction RMSE for the untransformed input, whereas the transformed rows give the RMSE between the original predictions and the transformed-back predictions. 
The rotation-induced errors are on the order of $10^{-4}$, and the reflection-induced errors remain around $10^{-3}$--$10^{-2}$, both well below the untransformed prediction errors. 
These results indicate that m-LSR-Net preserves the learned surface dynamics under rigid transformations of the input point cloud.

\begin{table}[htbp]
    \centering
    \caption{Geometric-consistency results for m-LSR-Net(TF) on the point-cloud Allen--Cahn equation with $T=5$. The ``Original'' row reports prediction RMSE for the untransformed input, while the transformed rows report the RMSE between the original predictions and the corresponding transformed-back predictions.}
    \begin{tabular}{c c c|cccc}
        \hline
        Transformation & Axis / Plane & Angle ($^\circ$) & $T$ & $2T$ & $3T$ & $4T$ \\
         \hline
        None & Original  & 0
        & $5.49\times10^{-2}$
        & $1.03\times10^{-1}$
        & $1.43\times10^{-1}$
        & $1.74\times10^{-1}$ \\
        \hline
        Rotation & $x$  & 60
        & $1.05\times10^{-4}$
        & $1.40\times10^{-4}$
        & $2.24\times10^{-4}$
        & $2.92\times10^{-4}$ \\

        Rotation & $y$  & 60
        & $1.38\times10^{-4}$
        & $1.81\times10^{-4}$
        & $2.79\times10^{-4}$
        & $3.99\times10^{-4}$ \\

        Rotation & $z$  & 60
        & $2.46\times10^{-4}$
        & $3.27\times10^{-4}$
        & $5.13\times10^{-4}$
        & $5.68\times10^{-4}$ \\
        \hline
        Reflection & $xy$ & --
        & $2.00\times10^{-3}$
        & $3.54\times10^{-3}$
        & $7.43\times10^{-3}$
        & $1.14\times10^{-2}$ \\

        Reflection & $yz$ & --
        & $1.68\times10^{-3}$
        & $2.49\times10^{-3}$
        & $5.93\times10^{-3}$
        & $1.01\times10^{-2}$ \\

        Reflection & $xz$ & --
        & $2.10\times10^{-3}$
        & $3.47\times10^{-3}$
        & $7.59\times10^{-3}$
        & $1.14\times10^{-2}$ \\
        \hline
    \end{tabular}
    \label{tab:ac_symmetry_rmse}
\end{table}

\section{Conclusion}
We have introduced LSR-Net, a long-short-range operator-learning framework for predicting pattern-forming dynamics on regular grids, spherical point clouds, and point clouds sampled from general manifolds. 
The main idea is to separate the learned evolution operator into a smooth nonlocal component and a geometry-dependent local component. 
The long-range component is represented by an SOE-based Fourier multiplier, which provides a compact parameterization of nonlocal interactions and can be evaluated efficiently through FFT-based implementations. 
The short-range component is chosen according to the data representation: standard local convolutions on planar grids, DISCO-based operators on the sphere, and DeltaConv-based local processing on general-manifold point clouds.

This decomposition leads to a modular architecture that can be adapted across different geometries without changing the basic LR--SR pipeline. 
For regular grids, the resulting p-LSR-Net substantially improves long-horizon prediction for the Allen--Cahn and Cahn--Hilliard equations. 
For spherical point clouds, s-LSR-Net substantially outperforms SFNO on the Allen--Cahn, Schnakenberg, and Turing systems. Finally
for Allen-Cahn dynamics on a blob-shaped manifold, m-LSR-Net achieves lower multi-step prediction errors than both GINO and the SR-only SR-Net(DeltaConv) baseline. 
The geometric consistency tests further show that the spherical and general-manifold implementations produce compatible predictions under rotations and reflections, despite the use of interpolation, Gaussian gridding, and autoregressive rollout.

Overall, the results support the use of the SOE-based LR module as a stable nonlocal component while leaving local and discretization-specific effects to geometry-informed SR operators. 
This separation is useful both computationally and architecturally: it allows the same long-range parameterization to be reused across data representations, while the local operator can be selected to respect the structure of the underlying geometry.

Several extensions remain open. 
The present work focuses on isotropic LR kernels and on pattern-forming systems over two-dimensional surfaces or planar domains. 
For higher-dimensional problems, alternatives to direct FFT-based realization, such as random Fourier features~\cite{rahimi2007random}, low-rank decompositions~\cite{march2017far}, and slicing approaches~\cite{hertrich2024fast1,hertrich2024fast2}, may provide more scalable long-range implementations. 
The SOE ansatz can also be extended in principle to anisotropic settings~\cite{Jiang2015}, which would be important for direction-dependent transport and heterogeneous media. 
The source code for LSR-Net is available on GitHub at https://github.com/qhou637/LSR-Geo-Net.

\section*{Acknowledgement}
The authors would like to acknowledge financial support from the Natural Science Foundation of China (Grant No. 12201146) and the Natural Science Foundation of Guangdong Province (Grant No. 2023A1515012197).

\appendix

\section{The Effect of the Tikhonov-Type Frequency-Reweighting Factor}
\label{app:Tik}

This appendix summarizes the effect of the Tikhonov-type frequency-reweighting factor used in the LR module,
$w_i(\mathbf{k}) = k^2/(k^2 + h_i^2)$,
where $h_i$ is a positive trainable parameter associated with the $i$-th LR channel (see Eq.~\eqref{eq:filter_fi}). 
We report TF-trained results for two spherical benchmarks: Allen--Cahn and Schnakenberg. 
The corresponding model settings are listed in~\ref{app:Model parameter settings}.

\subsection{Allen--Cahn}

\begin{figure*}[htb!]  
    \centering
    \renewcommand{\arraystretch}{0.8}  
    \setlength{\tabcolsep}{1pt}        
    \scriptsize
\includegraphics[width=0.8\textwidth]{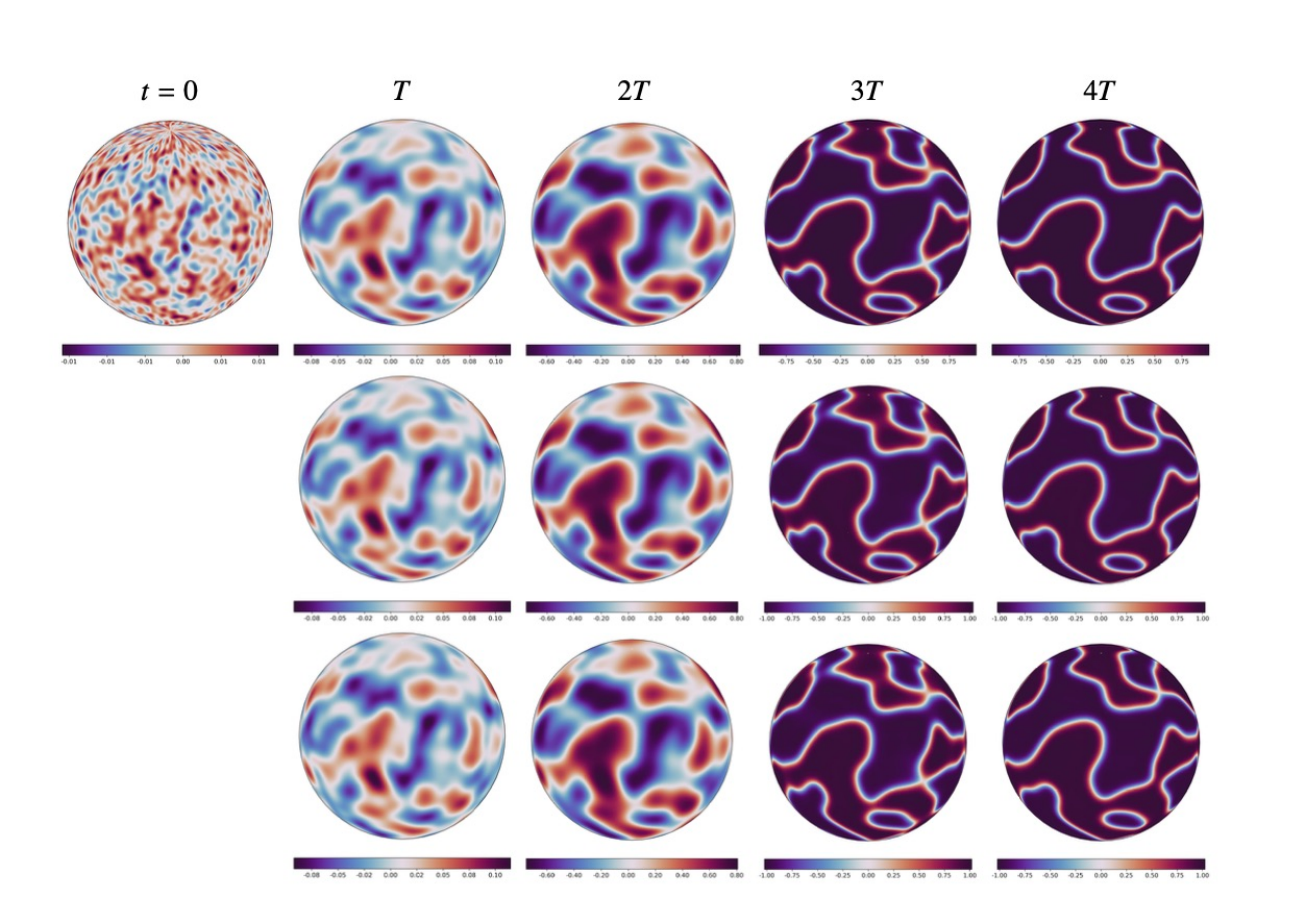}
    \caption{
    Allen--Cahn predictions on the sphere.
	    The first row shows the input and reference solutions at times $T$, $2T$, $3T$, and $4T$ ($T=3$).
	    The second and third rows show s-LSR-Net(TF) predictions without and with the Tikhonov-type frequency-reweighting factor, respectively.
	    }
    \label{fig:ac_infer_Tik_compare}
\end{figure*}

Fig.~\ref{fig:ac_infer_Tik_compare} compares predictions with and without reweighting.

\begin{table}[htbp]
    \centering
	    \caption{RMSE of s-LSR-Net(TF) with and without the Tikhonov-type frequency-reweighting factor on Allen--Cahn ($T=3$).}
    \begin{tabular}{c|cccc}
        \hline
        Method & $T$ & $2T$ & $3T$ & $4T$ \\
        \hline
	        Without reweighting   & 0.0004 & 0.0006 & 0.0011 & 0.0018 \\
	        With reweighting      & 0.0004 & 0.0005 & 0.0010 & 0.0017 \\
        \hline
    \end{tabular}
    \label{tab:ac_sph_rmse_Tik}
\end{table}
Table~\ref{tab:ac_sph_rmse_Tik} confirms that reweighting changes the Allen--Cahn errors only marginally, with differences in RMSE on the order of $10^{-4}$ up to $4T$.

\subsection{Schnakenberg System}
\begin{figure*}[htb!]  
    \centering
    \renewcommand{\arraystretch}{0.8}  
    \setlength{\tabcolsep}{1pt}        
    \scriptsize
\includegraphics[width=0.9\textwidth]{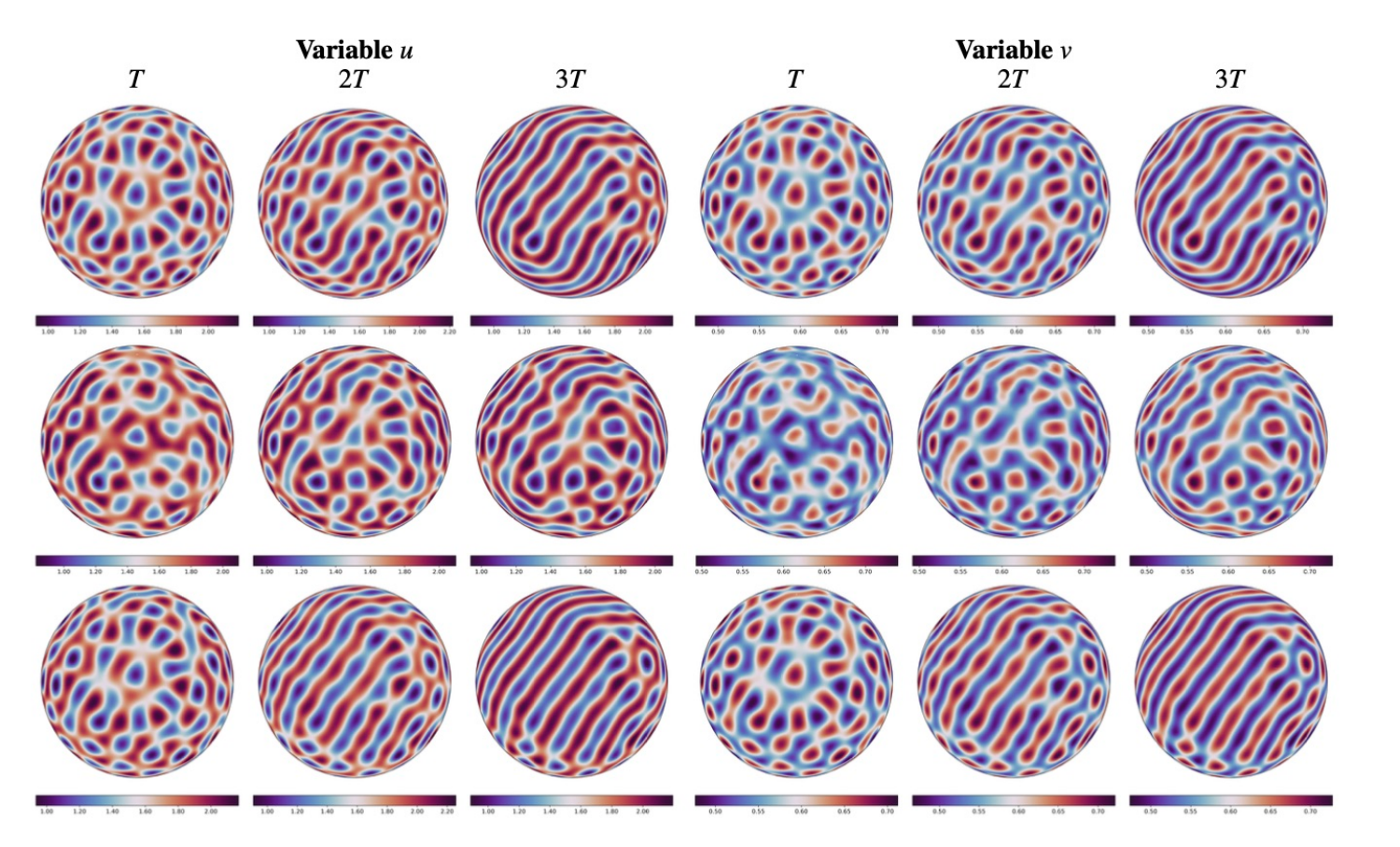}
    \caption{
	        Schnakenberg predictions for variables $u$ (left) and $v$ (right) at times $T$, $2T$, and $3T$ ($T=260$). Rows correspond to the reference solution, s-LSR-Net(TF) without reweighting, and s-LSR-Net(TF) with reweighting.}
    \label{fig:sch_infer_uv_Tik}   
\end{figure*}

For the Schnakenberg system, the reweighting factor has a clearer effect, as shown in Fig.~\ref{fig:sch_infer_uv_Tik} and Table~\ref{tab:schna_sph_rmse_Tik}.

\begin{table}[htbp]
    \centering
	    \caption{RMSE of s-LSR-Net(TF) with and without the Tikhonov-type frequency-reweighting factor for the Schnakenberg system ($T=260$).}
    \begin{tabular}{c|ccc}
        \hline
        Method            & $T$      & $2T$      & $3T$      \\
        \hline
	        Without reweighting     & 0.0048        & 0.0076        & 0.0098        \\
	        With reweighting      & 0.0004        & 0.0015        & 0.0039        \\
        \hline
    \end{tabular}
    \label{tab:schna_sph_rmse_Tik}
\end{table}
Table~\ref{tab:schna_sph_rmse_Tik} confirms the improvement quantitatively.
The RMSE is reduced by more than an order of magnitude at $T$, and the advantage persists through $3T$.
These results suggest that the frequency-reweighting is more important for the coupled Schnakenberg dynamics than for the single-field Allen--Cahn problem.

\section{LSR-U-Net}
\label{app:LSR-U-Net}
This appendix presents more details about the spherical U-Net variant used by s-LSR-Net. 
The LR module is inserted at the bottleneck of a U-Net encoder--decoder; DISCO blocks provide local spherical processing, and skip connections preserve fine-scale information.
Fig.~\ref{fig:arch_NN_unet} illustrates the resulting architecture.
Each encoder stage applies a DISCO block and then resamples the spherical grid (with details described below).
This halves the spatial resolution and doubles the channel width.
The number of encoder--decoder stages is treated as a tunable hyperparameter.

Let the input spherical grid be
$\{\theta_i\}_{i=0}^{N_{\mathrm{in}}-1}$ with $\theta_i \in [0,\pi]$, and let
$\{\theta'_j\}_{j=0}^{N_{\mathrm{out}}-1}$ with $\theta'_j \in [0,\pi]$ denote the output grid.
When input and output grids use different resolutions, their latitude nodes generally do not coincide:
\begin{equation}
\theta_i \neq \theta'_j.
\end{equation}
Resampling is therefore defined by interpolation of the input field at the output locations:
\begin{equation}
f_{\mathrm{out}}(\theta'_j, \phi)
\approx
f_{\mathrm{in}}(\theta'_j, \phi).
\end{equation}
For any $\theta'_j$, we find neighboring input points
$\theta_k \le \theta'_j \le \theta_{k+1}$ and compute
\begin{equation}
f_{\mathrm{out}}(\theta'_j, \phi)
=
(1 - w) f_{\mathrm{in}}(\theta_k, \phi)
+
w f_{\mathrm{in}}(\theta_{k+1}, \phi),
\end{equation}
where
\begin{equation}
w =
\frac{\theta'_j - \theta_k}
     {\theta_{k+1} - \theta_k}.
\end{equation}
At the bottleneck, the DISCO-based SR component and SOE-based LR component are applied in parallel.
The decoder mirrors the encoder through spherical upsampling, progressively recovering spatial resolution while reducing channel width.
A final $1\times 1$ convolution maps the reconstructed features to the physical output field.

Finally, we evaluate this s-LSR-Net architecture design on the Allen--Cahn, Schnakenberg, and Turing systems.
Table~\ref{tab:spher_unet_rmse_comparison} compares s-LSR-Net(AR) with and without the U-Net hierarchy, which clearly validates its improved accuracy on spherical data.
\begin{table}[htbp]
    \centering
    \caption{RMSE of s-LSR-Net(AR) on Allen--Cahn ($T=3$), Schnakenberg ($T=260$), and Turing ($T=130$) systems with LSR-U-Net and base LSR-Net architectures, respectively.}
    \begin{tabular}{c|c|cccc}
        \hline
        PDEs & Method & $T$ & $2T$ & $3T$ & $4T$ \\
        \hline
        Allen--Cahn 
            & LSR-U-Net          & 0.0005 & 0.0004 & 0.0006 & 0.0010 \\
            & base LSR-Net  & 0.8665 & 0.9579 & 0.9904 & 0.9946 \\
        \hline
        Schnakenberg 
            & LSR-U-Net         & 0.0017 & 0.0043 & 0.0076 & -- \\
            & base LSR-Net   & 0.0121 & 0.0153 & 0.0178 & -- \\
        \hline
        Turing 
            & LSR-U-Net         & 0.0231 & 0.1689 & 0.2116 & -- \\
            & base LSR-Net   & 0.1560 & 0.3245 & 0.3678 & -- \\
        \hline
    \end{tabular}
    \label{tab:spher_unet_rmse_comparison}
\end{table}

\section{Geometry-Specific SR Operators}
\label{app:sr-operators}

This appendix summarizes the geometry-specific local operators and the GNO operator used in the main text. 
All these SR operators are adopted from existing work and are included only to make the notations in Section~\ref{sec:3} self-contained.

\subsection{DISCO Convolution on the Sphere}

For point clouds on the sphere, the SR operator is taken to be the discrete--continuous (DISCO) convolution~\cite{ocampo2022disco}. Given a spherical signal $\phi$ and a continuous kernel $\mathcal{K}_{\mathrm{SR}}^{\mathrm{sph}}$, the DISCO convolution is approximated by
\begin{equation}
    (\phi \,\overset{\star}{ }\, \mathcal{K}_{\mathrm{SR}}^{\mathrm{sph}})(R)
    \;\approx\;
    \sum_{i} \phi[\omega_i]\, \mathcal{K}_{\mathrm{SR}}^{\mathrm{sph}}({R^{-1}\omega_i})\, q(\omega_i),
    \qquad R \in \mathrm{SO}(3),
\end{equation}
where $\{\omega_i\}$ are quadrature samples on the sphere with weights $q(\omega_i)$. Because the kernel is defined continuously on $\mathbb{S}^2$ and the rotation acts continuously on the domain, the resulting operator is equivariant with respect to rotations, up to quadrature error:
\begin{equation}
    ((Q\phi) \,\overset{\star}{ }\,\mathcal{K}_{\mathrm{SR}}^{\mathrm{sph}} )(R)
    \;=\;
    (Q (\phi \,\overset{\star}{ }\, \mathcal{K}_{\mathrm{SR}}^{\mathrm{sph}}))( R),
    \qquad Q \in \mathrm{SO}(3).
\end{equation}

\subsection{DeltaConv on General Manifolds}

For point clouds on general manifolds, the SR operator is taken to be DeltaConv~\cite{wiersma2022deltaconv}. Let $\mathcal{M}$ be a smooth surface and $\{p_i\}_{i=1}^N \subset \mathcal{M}$ a set of point samples. Scalar fields $\phi : \mathcal{M} \to \mathbb{R}$ are defined on the surface, while vector fields $v : \mathcal{M} \to T\mathcal{M}$ assign to each point $p \in \mathcal{M}$ a tangent vector $v(p) \in T_p\mathcal{M}$. Let $G$ and $D$ denote discrete gradient and divergence operators, and let $J$ denote the $\pi/2$ rotation on tangent vectors. The discrete Hodge Laplacian is
\begin{equation}
    L = - (G D - J G D J),
\end{equation}
which provides a diffusion mechanism for tangent-vector features consistent with the manifold geometry.

DeltaConv maintains scalar features $\phi_i$ and tangent-vector features $v_i \in T_{p_i}\mathcal{M}$ at each point $p_i$. The vector features are updated by
\begin{equation}
v_i' = h_{\Theta_0}\Big(v_i,\,(G\phi)_i,\,(Lv)_i\Big),
\end{equation}
where $(G\phi)_i$ is the intrinsic gradient of the scalar field and $(Lv)_i$ diffuses vector features through the Hodge Laplacian. The updated scalar features are then obtained by
\begin{equation}
\phi_i' = h_{\Theta_1}\Big(\phi_i,\,(Dv')_i,\ (-DJv')_i,\ \|v_i'\|\Big)
          + \max_{j\in\mathcal{N}_i} h_{\Theta_2}(\phi_j).
\end{equation}
This coupled scalar--vector update enables intrinsic anisotropic local processing on the manifold.

\subsection{GNO on General Manifolds}

In the general-manifold LR branch, GNO maps point-cloud features to a latent regular grid. 
Given an input point cloud $\{\mathbf{x}_i^{\mathrm{in}}\}_{i=1}^N \subset S$ with associated values $\{\phi_i\}_{i=1}^N$, the GNO encoder defines features on a uniform latent grid $\{\mathbf{x}_\ell^{\mathrm{grid}}\}_{\ell=1}^M \subset D$.
This can be interpreted as a discretized local integral operator.
For each grid point $\mathbf{x}_\ell^{\mathrm{grid}}$, information is aggregated from nearby samples within
\[
B_r(\mathbf{x}_\ell^{\mathrm{grid}}) := \{\mathbf{y} \in S : \|\mathbf{y} - \mathbf{x}_\ell^{\mathrm{grid}}\| \le r\},
\]
leading to
\begin{equation}
\label{eq:gno_lifting}
v_0(\mathbf{x}_\ell^{\mathrm{grid}})
\approx
\sum_{i=1}^{N}
\kappa(\mathbf{x}_\ell^{\mathrm{grid}}, \mathbf{x}_i^{\mathrm{in}})\, 
\phi_i \, \mu(\mathbf{x}_i^{\mathrm{in}}),
\end{equation}
where $\kappa(\cdot,\cdot)$ is a learnable kernel depending on the relative 
geometry, and $\mu(\mathbf{x}_i^{\mathrm{in}})$ denotes the Riemannian quadrature weights.
The structured latent grid is then used by the LR module for nonlocal feature propagation.


\section{Turing System on Spherical Manifolds}
\label{app:Turing}

We also consider a two-species Turing system
\cite{Stewart1998,Turing1952} on a spherical surface $\Gamma$:

\begin{equation}
\frac{\partial u}{\partial t} = \delta_u \Delta_\Gamma u
+ \alpha u \left( 1 - \tau_1 v^2 \right) + v \left( 1 - \tau_2 u \right),
\end{equation}

\begin{equation}
\frac{\partial v}{\partial t} = \delta_v \Delta_\Gamma v
+ \beta v \left( 1 + u v \right) + u \left( \gamma + \tau_2 v \right),
\end{equation}
where $u$ and $v$ denote the concentrations of the two species. Depending on
the choice of parameters, solutions can form Turing patterns, including spots
and stripes. On the spherical surface, the parameters are chosen as:
\[
\delta_v = 1\times 10^{-3}, \quad
\delta_u = 0.516\,\delta_v, \quad
\alpha = 0.899, \quad
\beta = -0.91, \quad
\gamma = -0.899, \quad
\tau_1 = 0.02, \quad
\tau_2 = 0.2.
\]

The spherical Turing system provides a multi-species benchmark for comparing AR and TF training within the same s-LSR-Net architecture. 
The s-LSR-Net model uses embedding channels $=7$, Fourier channels $=3$, downsampling factor $=3$, kernel size $=5$, and multi-step rollout length $=3$.

\begin{table}[htbp]
    \centering
    \caption{RMSE of s-LSR-Net and SFNO on the Turing system ($T=130$).}
    \begin{tabular}{c|ccc}
        \hline
        Method            & $T$     & $2T$     & $3T$     \\
        \hline
        SFNO              & 0.0493       & 0.6747       & 0.7206       \\
        s-LSR-Net(AR)      & 0.0231       & \textbf{0.1689}       & \textbf{0.2116}       \\
        s-LSR-Net(TF)      & \textbf{0.0114}       & 0.1920       & 0.2651       \\
        \hline
    \end{tabular}
    \label{tab:turing_rmse}
\end{table}

\begin{figure*}[htb!]  
    \centering
    \renewcommand{\arraystretch}{0.8}  
    \setlength{\tabcolsep}{1pt}        
    \scriptsize
\includegraphics[width=0.9\textwidth]{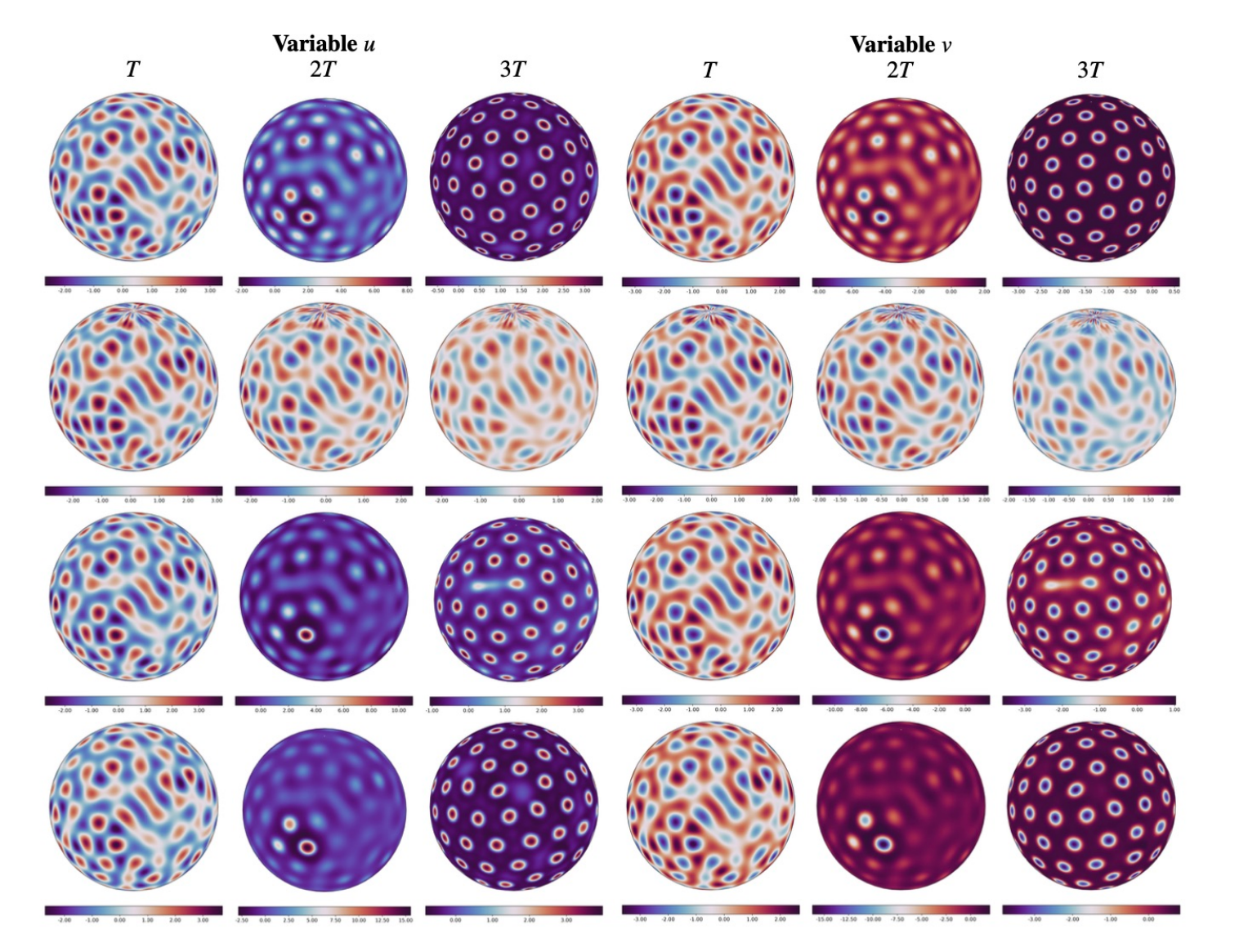}
	    \caption{
	        Predictions of the Turing system for variables \(u\) (left) and \(v\) (right) at times \(T\), \(2T\), and \(3T\) with \(T=130\). 
	        Rows correspond to the reference solution, SFNO, s-LSR-Net(AR), and s-LSR-Net(TF), respectively.
	    }
	    \label{fig:turing_infer_uv}
\end{figure*}

Fig.~\ref{fig:turing_infer_uv} compares both species at $T$, $2T$, and $3T$ with $T=130$.
Both s-LSR-Net variants preserve the main pattern structures more accurately than SFNO.
TF gives the smallest short-horizon error, while AR gives slightly smaller errors at later horizons, as documented in Table~\ref{tab:turing_rmse}.


\section{Data Preparation}
\label{app:Data Preparation}
\subsection{Euclidean 2D Plane}
For the two-dimensional Euclidean setting, Allen--Cahn simulations are performed on a square domain $\Omega = [0,1]^2$ using a uniform spatial grid of resolution $128 \times 128$. The initial conditions are sampled from a Gaussian random field (GRF). The Cahn--Hilliard simulations are conducted on a square domain $\Omega = [-1,1]^2$ with the same spatial resolution.
Unless otherwise specified, the parameters are set to $M = 1$, $\kappa = 8$, and $W = 2$. The initial conditions are sampled from a Gaussian random field with an added constant bias, giving mean value $0.5$ and preserving mass conservation.
The model is trained using solution data up to a final time $T = 7$, and is subsequently employed to autoregressively predict future states at extended time horizons $2T, 3T, \dots, 7T$.
For each PDE, we generate $N=500$ early-stage snapshots under diverse initial conditions and construct training, validation, and test splits accordingly.

\subsection{Point Clouds on a Sphere}
All datasets are generated on the unit sphere $\Gamma$ using spectral discretization based on spherical harmonics.
Spatial differential operators are evaluated via the Laplace--Beltrami operator $\Delta_\Gamma$, and time integration is performed in the spectral domain.
The spherical surface is discretized using a latitude--longitude grid with $n_{\mathrm{lat}} = 256$ and $n_{\mathrm{lon}} = 512$.
Spherical harmonic expansions are truncated at $l_{\max} = m_{\max} = \lceil n_{\mathrm{lon}}/2 \rceil$.
At each prescribed output time, the spectral coefficients are transformed back to the physical grid to form point-cloud data on the sphere.

The Allen--Cahn equation on the sphere uses the same governing dynamics as in the Euclidean setting; the difference lies in the surface discretization and in the use of the Laplace--Beltrami operator.
Initial conditions are sampled as Gaussian random fields in the spectral domain and transformed to the grid representation, constrained within $[-1,1]$ for numerical stability.
The interface parameter is set to $\epsilon = 10^{-3}$.
Time integration is carried out using an explicit Adams--Bashforth scheme with a time step $\Delta t = 10^{-3}$ over a total simulation time $T=12$.
Snapshots are recorded at uniform time intervals.

For the Schnakenberg system, parameters are chosen as $D_u = 0.001$, $D_v = 0.02$, $a = 0.1$, and $b = 1.5$ to generate stripe patterns.
Initial conditions are generated as Gaussian random fields in the spectral domain for both species.
The system is evolved for a sufficiently long time horizon to allow pattern formation, and solution snapshots are recorded at regular intervals.
Each dataset sample consists of paired $(u,v)$ fields represented on the spherical grid.

Details of Turing system data generation are provided in~\ref{app:Turing}.
For each PDE, $N=400$ early-stage snapshots are generated under diverse initial conditions, and training, validation, and test splits are constructed accordingly.

\subsection{Point Clouds on General Manifolds}
For the Allen--Cahn equation posed on a smooth, closed surface
$\Gamma \subset \mathbb{R}^3$,
 $\Gamma$ is chosen as a smooth \emph{blob-shaped} surface,
characterized by nontrivial curvature and a geometry that cannot be covered by a single global parametrization.
Rather than relying on structured grids or intrinsic coordinates, the surface is represented by an unstructured point cloud
$\mathcal{P}=\{x_i\}_{i=1}^{N} \subset \Gamma$, where each point
$x_i=(x_i,y_i,z_i)$ corresponds to a location on the embedded surface. This representation avoids coordinate singularities and supports general curved geometries.

To generate high-fidelity numerical data on general manifolds, we employ
the \texttt{Surfacefun} framework~\cite{fortunato2024high}, which provides
a high-order, efficient solver for PDEs posed on smooth surfaces
embedded in $\mathbb{R}^3$. The surface is discretized into multiple
overlapping spectral patches, each with local coordinates and an induced
metric from the embedding. Intrinsic differential operators, including the
Laplace--Beltrami operator, are assembled through this patch-based
discretization, enabling accurate evaluation of surface diffusion
without relying on global coordinates.

For these simulations, we use a smooth blob-shaped surface generated by \texttt{Surfacefun}.
The resolution parameter is \(n=20\), which defines the number of spectral patches and
resolves both the geometric features of the surface and the nonlinear Allen--Cahn dynamics.
The diffusion coefficient is set to
$\delta = 10^{-3}$, with a time step of $\Delta t = 10^{-2}$.

\section{Model Parameter Settings}
\label{app:Model parameter settings}

For the Euclidean two-dimensional plane, the model parameter settings of base p-LSR-Net and FNO for the Allen--Cahn and Cahn--Hilliard equations are summarized in Tables~\ref{tab:2D_model_settings} and~\ref{tab:2D_model_settings_fno}, respectively.

\begin{table}[htbp]
\centering
\caption{Euclidean 2D plane: Model settings for inferring the Allen--Cahn and Cahn--Hilliard equations using p-LSR-Net.}
\begin{tabular}{lcc}
\hline
Parameter & Allen--Cahn & Cahn--Hilliard \\
\hline
Pre-training epochs      & 30   & 30      \\
Optimizer                & Adam & Adam    \\
Learning rate             & 0.001 & 0.001  \\
Weight initialization    & Xavier & Xavier \\
Training kernel size     & 7    & 7       \\
Number of training layers & 3    & 3       \\
Training channels (SR module) & 7 & 7     \\
Training channels (LR module) & 1 & 1     \\
Gradient coefficient     & 0    & 0.3     \\
\hline
\end{tabular}
\label{tab:2D_model_settings}
\end{table}

\begin{table}[htbp]
\centering
\caption{Model parameter settings for FNO in the Euclidean 2D plane.}
\begin{tabular}{lcc}
\hline
Parameter & Allen--Cahn & Cahn--Hilliard \\
\hline
Number of layers     & 4   & 4    \\
Fourier modes        & 32  & 6    \\
Hidden channels      & 10  & 10   \\
Optimizer            & AdamW & AdamW \\
Learning rate        & $8\times 10^{-3}$ & $8\times 10^{-3}$ \\
\hline
\end{tabular}
\label{tab:2D_model_settings_fno}
\end{table}
\begin{itemize}
    \item Gaussian gridding (mollification) is also used in the regular 2D grid setting. Its impact is relatively mild for the second-order Allen--Cahn equation, but it becomes more beneficial for the fourth-order Cahn--Hilliard equation.
 \item In the short-range (SR) module, for the Allen--Cahn equation, the SR component consists solely of convolutional layers. For the Cahn--Hilliard equation, the SR component is further augmented with downsampling and upsampling operations to capture high-order local features. Specifically, downsampling is performed via a strided convolution with stride $5$, while upsampling is implemented using a transposed convolution with the same stride.
\end{itemize}
For point clouds on a sphere, model settings for inferring the Allen--Cahn, Schnakenberg, and Turing systems using s-LSR-Net and SFNO are provided in Tables~\ref{tab:sphere_model_settings} and~\ref{tab:sphere_model_settings_sfno}.

\begin{table}[htbp]
\centering
\caption{Model settings for inferring the Allen--Cahn, Schnakenberg, and Turing systems on the sphere using s-LSR-Net.}
\begin{tabular}{lccc}
\hline
Parameter & Allen--Cahn & Schnakenberg & Turing \\
\hline
Epochs               & 50    & 50    & 50    \\
Optimizer            & Adam  & Adam  & Adam  \\
Learning rate        & $1\times10^{-3}$ & $1\times10^{-3}$ & $1\times10^{-3}$ \\
Layers               & 4     & 4     & 4     \\
Feature embedding dimension & 7     & 7     & 7     \\
LR Fourier channels   & 2     & 2     & 3     \\
Number of downsampling stages & 4     & 3     & 3     \\
Sphere kernel shape  & $(3,4)$ & $(5,6)$ & $(5,6)$ \\
\hline
\end{tabular}
\label{tab:sphere_model_settings}
\end{table}

\begin{itemize}
    \item \textbf{Feature embedding dimension:} Number of feature channels at each spherical grid point in the short-range (SR) module; controls the width of learned representations.
    \item \textbf{Number of downsampling stages:} Number of encoder-decoder stages in the U-Net; each stage halves the spatial resolution and typically increases the number of channels.
    \item \textbf{LR Fourier channels:} Number of channels used in the long-range (LR) module (NUFFT/Fourier) to capture global interactions on the sphere.
    \item \textbf{Sphere kernel shape:} Angular size of the convolution kernel in the SR DISCO block, given as $(\theta, \phi)$; determines the local receptive field on the spherical grid.
\end{itemize}

\begin{table}[htbp]
\centering
\caption{Model settings for inferring the Allen--Cahn, Schnakenberg, and Turing systems on the sphere using SFNO.}
\begin{tabular}{lccc}
\hline
Parameter & Allen--Cahn & Schnakenberg & Turing \\
\hline
Epochs                    & 50    & 50    & 50    \\
Optimizer                  & Adam  & Adam  & Adam  \\
Learning rate              & $1\times10^{-3}$ & $1\times10^{-3}$ & $1\times10^{-3}$ \\
Number of layers           & 4     & 4     & 4     \\
Feature embedding dimension & 7    & 7     & 7     \\
Local kernel shape         & $(3,4)$ & $(5,6)$ & $(5,6)$ \\
Use MLP in SFNO blocks     & Yes   & Yes   & Yes   \\
Big skip connection        & True  & True  & True  \\
\hline
\end{tabular}
\label{tab:sphere_model_settings_sfno}
\end{table}

\begin{itemize}
    \item \textbf{Feature embedding dimension:} Number of feature channels at each spherical grid point inside each SFNO block.
    \item \textbf{Number of layers:} Total number of SFNO blocks stacked sequentially; each block alternates between global spectral convolution and local convolution.
    \item \textbf{Local kernel shape:} Size of the short-range convolution kernel on the sphere, given as $(\theta, \phi)$.
    \item \textbf{Use MLP in SFNO blocks:} Whether a small MLP is applied inside each block for feature mixing.
    \item \textbf{Big skip connection:} Whether a residual connection from input to output is added across the entire SFNO network.
\end{itemize}

For point clouds on general manifolds, the Allen--Cahn model settings are provided for SR-Net(DeltaConv), GINO, and m-LSR-Net in Tables~\ref{tab:general_deltaconv_model_settings}, \ref{tab:general_gino_model_settings}, and~\ref{tab:general_model_settings}, respectively.

\begin{table}[htbp]
\centering
\caption{Model settings for inferring the Allen--Cahn system on general manifolds using SR-Net(DeltaConv).}
\begin{tabular}{lc}
\hline
Parameter & Value \\
\hline
Epochs                        & 30 \\
Optimizer                     & Adam \\
Learning rate                  & $1\times10^{-3}$ \\
Weight decay                   & $0$ \\
Forward step                   & 4 \\
\hline
Input channels                 & 1 \\
DeltaConv convolution channels & [16, 32, 64] \\
MLP depth                      & 3 \\
Embedding size                 & 10 \\
Number of neighbors            & 30 \\
Gradient regularizer           & 0.001 \\
Gradient kernel width          & 1.0 \\
\hline
\end{tabular}
\label{tab:general_deltaconv_model_settings}
\end{table}

\begin{itemize}
    \item \textbf{Forward step:} Number of time steps predicted per autoregressive step.
    \item \textbf{DeltaConv convolution channels:} Number of channels in each convolutional layer, controlling the model capacity for local interactions.
    \item \textbf{MLP depth:} Number of layers in the pointwise MLPs applied after the DeltaConv operations.
    \item \textbf{Embedding size:} Dimension of the latent representation learned for each point.
    \item \textbf{Number of neighbors:} Number of neighboring points considered for each DeltaConv operation.
    \item \textbf{Gradient regularizer:} Strength of the regularization term applied to gradient predictions, improving stability.
    \item \textbf{Gradient kernel width:} Width parameter for the gradient regularization kernel.
    \item \textbf{Input channels:} Dimension of the input features for each point in the point cloud.
\end{itemize}

\begin{table}[htbp]
\centering
\caption{Model settings for inferring the Allen--Cahn system on general manifolds using GINO.}
\begin{tabular}{lc}
\hline
Parameter & Value \\
\hline
Epochs                        & 25 \\
Optimizer                     & Adam \\
Learning rate                  & $1\times10^{-3}$ \\
Weight decay                  & $1\times10^{-5}$ \\
Forward step                   & 4 \\

Number of FNO layers           & 3 \\
FNO hidden channels            & 32 \\
FNO Fourier modes              & 16 \\

GINO projection channels       & 7 \\
GNO embedding channels         & 32 \\
GNO coordinate dimension       & 3 \\
GNO radius                     & 0.2 \\

GINO input channel MLP         & [64, 64, 64] \\
GINO output channel MLP        & [128, 128] \\

NUFFT latent grid size         & $10 \times 10 \times 10$ \\

\hline
\end{tabular}
\label{tab:general_gino_model_settings}
\end{table}

\begin{itemize}
    \item \textbf{Forward step:} Number of time steps predicted per autoregressive step.
    \item \textbf{FNO layers, hidden channels, and Fourier modes:} Control the depth and spectral resolution of the global latent dynamics modeled by the Fourier Neural Operator.
    \item \textbf{GINO projection channels:} Dimension of the lifted latent representation connecting point clouds and the latent NUFFT grid.
    \item \textbf{GNO embedding channels and radius:} Define the local geometric operator for aggregating neighborhood information on the manifold point cloud.
    \item \textbf{Input/output channel MLPs:} Nonlinear feature transformations applied before and after the geometric neural operator.
    \item \textbf{Latent grid size:} Resolution of the NUFFT-based structured grid used for global convolution.
\end{itemize}

\begin{table}[htbp]
\centering
\caption{Model settings for inferring the Allen--Cahn system on general manifolds using m-LSR-Net.}
\begin{tabular}{lc}
\hline
Parameter & Value \\
\hline
Epochs                        & 25 \\
Optimizer                     & Adam \\
Learning rate                  & $1\times10^{-3}$ \\
Forward step                   & 4 \\
Number of LSR blocks           & 3 \\
DeltaConv channels             & [64, 128, 256] \\
DeltaConv embedding size       & 64 \\
DeltaConv num neighbors        & 30 \\
DeltaConv MLP depth            & 2 \\
GINO projection channels       & 7 \\
NUFFT FNO channels             & 2 \\
Grid size (NUFFT mesh)         & 10 \\
GNO hidden channels            & 32 \\
\hline
\end{tabular}
\label{tab:general_model_settings}
\end{table}

\begin{itemize}
    \item \textbf{Forward step:} Number of time steps predicted per autoregressive step.
    \item \textbf{DeltaConv channels / embedding size / num neighbors:} Control short-range local feature extraction at each point.
    \item \textbf{GINO projection channels / NUFFT FNO channels:} Control long-range feature representation on the point cloud via NUFFT.
    \item \textbf{Number of LSR blocks:} Number of LSR blocks stacked sequentially.
    \item \textbf{Grid size:} Number of points in each dimension of the latent NUFFT grid.
\end{itemize}

\end{document}